\PassOptionsToPackage{table}{xcolor}
\documentclass[]{nicetemplate}

\usepackage{enumitem}
\usepackage{amssymb}
\usepackage{mathtools}
\usepackage{amsthm}

\usepackage{graphicx}
\usepackage{booktabs}
\usepackage{array}
\usepackage{url}
\usepackage{xurl}
\usepackage{placeins}
\usepackage{needspace}
\usepackage[table]{xcolor}
\usepackage{tabularx}

\definecolor{checkgreen}{HTML}{228B61}
\definecolor{crossred}{HTML}{C45454}
\definecolor{highlightblue}{HTML}{EEF5FC}
\definecolor{envc}{HTML}{21806D}\definecolor{envbg}{HTML}{E3F2EE}
\definecolor{taskc}{HTML}{6A4C9C}\definecolor{taskbg}{HTML}{EFEAF7}
\newcommand{\colgap}{\hspace{14pt}}
\newcommand{\grouprow}[3]{%
  \rowcolor{#1bg}\multicolumn{3}{@{\hspace{4pt}}l}{\rule{0pt}{2.6ex}\textcolor{#1c}{\textbf{#2}}\quad\textit{#3}}\\[2pt]}

\newcommand{\dn}[1]{\textcolor{black!50}{#1}}   
\newcommand{\bandrow}[3]{\rowcolor{#1bg}\multicolumn{#3}{@{\hspace{4pt}}l}{\rule{0pt}{2.6ex}\textcolor{#1c}{\textbf{#2}}}\\[1pt]}
\newcommand{\tabyes}{\textcolor{checkgreen}{$\checkmark$}}
\newcommand{\tabno}{\textcolor{crossred}{$\times$}}
\newcolumntype{Y}{>{\centering\arraybackslash}X}
\makeatletter\def\input@path{{./}{../}}\makeatother
\graphicspath{{./}{../}}

\newcolumntype{L}[1]{>{\raggedright\arraybackslash}p{#1}}
\theoremstyle{plain}

\theoremstyle{definition}

\theoremstyle{remark}

\newcommand{\PluginName}{\textsc{Screen2Bench}}

\title{\PluginName: From Screen Recordings to Faithful and Verifiable Computer-Use Benchmarks}

\author[1]{Yash Maheshwari}
\author[1]{Kaustubh D. Dhole}
\author[1]{Suhana Bedi}
\author[1]{Nigam H. Shah}
\affiliation[1]{Stanford University}
\date{\today}
\correspondence{\email{\{yashmahe, kdhole, suhana, nigam\}@stanford.edu}}

\metadata[Code]{%
  \href    {https://github.com/som-shahlab/screen2bench}
    {som-shahlab/screen2bench}%
}

\abstract{
Evaluating computer-use agents in realistic environments is challenging, particularly in healthcare, where access to proprietary applications is often restricted. Screen recordings, often available through public tutorials and software demonstrations, offer a practical alternative for constructing computer-use benchmarks without direct access to these applications. We present \PluginName{}, a multi-agent orchestrator implemented as a Claude Code plugin that automatically constructs computer-use benchmarks from screen recordings by reconstructing interactive environments and generating evaluation tasks with executable verifiers. \PluginName{} coordinates specialized agents to reconstruct recorded interfaces, independently assess their fidelity, and validate task verifiers against correct and incorrect task executions. To systematically assess the quality of the resulting benchmarks, we evaluate environment fidelity and task quality using multiple metrics and introduce two measures: task preference, which assesses how well generated tasks cover and evaluate recorded workflows, and check strength, inspired by mutation testing in software engineering, which quantifies the extent to which task verifiers detect incorrectly entered values. We evaluate our approach across three simulated healthcare applications by comparing Claude Code with and without the plugin under matched conditions, using identical screen recordings, base models (Claude Opus 5.5, Claude Sonnet 5.5, or Claude Haiku 5.5), and model effort levels. With Claude Opus 5.5, \PluginName{} wins 71 of 72 metric-level comparisons, including all comparisons on environment fidelity, with one tie and no losses. It wins 67 of 72 with Claude Sonnet 5.5 and 64 of 72 with Claude Haiku 5.5. These findings demonstrate the potential of reconstructing interactive environments and verifiable evaluation tasks from screen recordings to support scalable and reproducible benchmarking of computer-use agents.
}

\keywords{Computer-Use Agents, Automated Benchmark Construction, Multi-Agent Systems}

\begin{document}
\maketitle

\section{Introduction}

As computer-use agents become capable of performing complex workflows across software applications~\citep{zhou2024webarena,xu2026theagentcompany}, evaluating their reliability requires benchmarks that combine realistic, interactive environments with evaluation tasks whose outcomes can be verified. Such benchmarks enable researchers to assess agent reliability and safety before and during real-world deployment~\citep{levy2026st}, while also supporting reinforcement learning (RL)~\citep{ICLR2025_c66e1fcc,Bai2026CVPR} and progressively challenging evaluations~\citep{nam2026evolving,yin2026constructing}.

Constructing these benchmarks remains challenging because existing methods often require direct access to the original applications~\citep{rawles2025androidworld,wei2026opencomputer,cao2026guigenesis}. This requirement is particularly restrictive in healthcare, where applications are often proprietary and access is limited~\citep{bedi2026healthadminbench}. Screen recordings of user workflows, on the other hand, are easier to obtain and often available through software demonstrations and tutorials, offering an alternative for reconstructing interactive environments and generating evaluation tasks without accessing the original applications. However, screen recordings capture only observed interfaces and interactions, rather than the underlying application logic or the conditions required to verify task completion.

Automatically constructing reliable benchmarks from screen recordings therefore presents two distinct challenges. First, reconstructed environments must faithfully reproduce the recorded interfaces and interactions. Second, generated evaluation tasks must include verifiers that reliably distinguish correct from incorrect outcomes. For example, a verifier that checks only a subset of required form fields may incorrectly mark an incomplete task as successful. In our experiments, Claude Code~\citep{anthropic2025claudecode}, a general-purpose coding agent, exhibited both limitations: its reconstructed environments omitted or altered interface content, and its task verifiers failed to check some required values. These limitations motivate systematic assessment of \textbf{environment fidelity} and \textbf{task quality} in automatically generated benchmarks.

To address these challenges, we make the following contributions:

\begin{itemize}
    \item We introduce \PluginName{}, a multi-agent orchestrator implemented as a Claude Code plugin that reconstructs interactive environments and generates verifiable evaluation tasks from screen recordings, without requiring access to the original applications. \PluginName{} coordinates specialized agents to assess the fidelity of the reconstructed environments and validate task verifiers using correct and incorrect task executions.

    \item We present a systematic evaluation framework comprising multiple measures for assessing \textbf{environment fidelity} and \textbf{task quality} of benchmarks constructed from screen recordings. Within this framework, we introduce two new measures of task quality: \textbf{task preference}, which assesses how well generated tasks cover and evaluate recorded workflows, and \textbf{check strength}, inspired by software testing, which measures whether task verifiers detect incorrectly entered values.

    \item We evaluate \PluginName{} on three healthcare applications from HealthAdminBench~\citep{bedi2026healthadminbench}, comparing Claude Code with and without the plugin using identical screen recordings, the same base model (Claude Opus 5.5, Claude Sonnet 5.5, or Claude Haiku 5.5), and the same effort level. With Claude Opus 5.5, \PluginName{} wins 71 of 72 metric-level comparisons, loses none, and ties 1, including all comparisons on environment fidelity. It wins 67 of 72 with Claude Sonnet 5.5 and 64 of 72 with Claude Haiku 5.5.
\end{itemize}

The remainder of the paper is organized as follows.~\Cref{sec:rel_work} reviews related work.~\Cref{sec:method} formalizes the benchmark construction problem and introduces \PluginName{}, including its multi-agent architecture and approach to environment reconstruction, task generation, and validation.~\Cref{sec:exp} presents our experimental setup and evaluation framework, and~\Cref{sec:results} presents the results.

\section{Related Work}\label{sec:rel_work}

\subsection{Dynamic Benchmarking for Computer Use}

Recent work has explored scalable methods for constructing interactive environments and evaluation tasks for computer-use agents. AndroidWorld~\cite{rawles2025androidworld} and WorkArena++~\cite{boisvert2024workarenaplusplus} use parameterized task templates and randomized environment configurations to generate reproducible tasks with programmatic success verification. Cua-Bench~\cite{qi2026cuabench} further supports configurable desktop environments with randomized application states and visual conditions. Moving beyond predefined templates, AgentSynth~\cite{xie2026agentsynth} and GTA~\cite{huang2026gta} use large language models (LLMs) to generate evaluation tasks and execution trajectories, while OpenComputer~\cite{wei2026opencomputer} jointly generates tasks and task validators across heterogeneous desktop applications. Other approaches focus on evaluation diversity and difficulty: BreakingWeb~\cite{yin2026constructing} introduces controlled environment perturbations, while TURTLE~\cite{nam2026evolving} iteratively generates challenging tasks based on agent performance. Although these methods improve the scalability of computer-use evaluation, they largely operate within existing environments and require access to the original applications. Consequently, they do not directly address benchmark construction when the original applications are inaccessible.

Recent approaches have begun addressing environment reconstruction. GUI-GENESIS~\cite{cao2026guigenesis} reconstructs lightweight interactive environments with verifiable rewards, while VeriEnv~\cite{chae2026verienv} uses coding agents to reconstruct websites from screenshots and generate evaluation tasks with executable validators. IWR-Bench~\cite{chen2026iwrbench} studies interactive webpage reconstruction from user interaction videos and additional website assets, evaluating both visual and functional fidelity. Trace2World~\cite{wang2026trace2world} reconstructs interactive environments from recorded interaction traces, using source-grounded behavioral models and independent workflow verification to improve reconstruction fidelity. REA~\cite{morluto2026rea} reconstructs API-based environments from recorded network traffic, while ShowUI-Aloha~\cite{zhang2026showuialoha} uses screen recordings to train agents to interact with existing desktops rather than construct evaluation environments.

These approaches demonstrate the potential of automated environment reconstruction, task and verifier generation, but differ in their input requirements and the capabilities they combine. In particular, reconstructing reliable benchmarks from screen recordings alone requires both faithful reproduction of recorded interfaces and interactions and evaluation tasks with verifiers that reject incorrect outcomes. As summarized in~\Cref{tab:related_work_comparison}, \PluginName{} combines these capabilities without requiring access to the original applications, their source code, accessibility trees, or recorded API traffic. It further incorporates independent interface review and explicit testing of task verifiers against correct and incorrect executions, addressing both environment fidelity and task quality within a unified benchmark construction pipeline.

\begin{table*}[t]
\centering
\footnotesize
\setlength{\tabcolsep}{3pt}
\renewcommand{\arraystretch}{1.2}

\begin{tabularx}{\textwidth}{
@{}>{\raggedright\arraybackslash}p{0.29\textwidth}
*{6}{Y}@{}
}
\toprule

& \multicolumn{1}{c}{\textbf{Input}}
& \multicolumn{2}{c}{\textbf{Construction}}
& \multicolumn{3}{c}{\textbf{Validation}} \\

\cmidrule(lr){2-2}
\cmidrule(lr){3-4}
\cmidrule(lr){5-7}

\textbf{Method}
& \shortstack{\textbf{Video}\\\textbf{Only}}
& \shortstack{\textbf{GUI}\\\textbf{Clone}}
& \shortstack{\textbf{Auto}\\\textbf{Tasks}}
& \shortstack{\textbf{Task}\\\textbf{Checks}}
& \shortstack{\textbf{Screen}\\\textbf{Critic}}
& \shortstack{\textbf{Negative}\\\textbf{Tests}} \\

\midrule

AndroidWorld~\cite{rawles2025androidworld}
& \tabno & \tabno & \tabyes & \tabyes & \tabno & \tabno \\

WorkArena++~\cite{boisvert2024workarenaplusplus}
& \tabno & \tabno & \tabyes & \tabyes & \tabno & \tabno \\

AgentSynth~\cite{xie2026agentsynth}
& \tabno & \tabno & \tabyes & \tabno & \tabno & \tabno \\

OpenComputer~\cite{wei2026opencomputer}
& \tabno & \tabno & \tabyes & \tabyes & \tabno & \tabno \\

GUI-GENESIS~\cite{cao2026guigenesis}
& \tabno & \tabyes & \tabno & \tabyes & \tabno & \tabno \\

IWR-Bench~\cite{chen2026iwrbench}
& \tabno & \tabyes & \tabno & \tabno & \tabno & \tabno \\

VeriEnv~\cite{chae2026verienv}
& \tabno & \tabyes & \tabyes & \tabyes & \tabno & \tabno \\

REA~\cite{morluto2026rea}
& \tabno & \tabno & \tabno & \tabno & \tabno & \tabno \\

Trace2World~\cite{wang2026trace2world}
& \tabno & \tabyes & \tabyes & \tabyes & \tabyes & \tabyes \\

\midrule
\rowcolor{highlightblue}
\textbf{\PluginName{} (Ours)}
& \tabyes & \tabyes & \tabyes
& \tabyes & \tabyes & \tabyes \\

\bottomrule
\end{tabularx}

\caption{
Comparison of computer-use benchmark construction approaches.
\textcolor{checkgreen}{$\checkmark$} indicates an explicitly
supported capability and \textcolor{crossred}{$\times$}
indicates a capability not reported as part of the method.
Video Only denotes construction from screen recordings
without additional application assets, action logs, or API
traces. GUI Clone denotes reconstruction of an interactive
graphical environment. Task Checks denotes executable
task-success validators. Screen Critic denotes independent
agent review of reconstructed interfaces during construction.
Negative Tests denotes explicit testing that task validators
reject incorrect executions.
}
\label{tab:related_work_comparison}
\end{table*}

\subsection{Evaluating Computer-Use Agents for Healthcare}

Healthcare presents a particularly challenging setting for computer-use evaluation, as clinical and administrative workflows involve complex interactions across electronic health record (EHR) systems, insurance portals, and other software applications. Recent benchmarks have established realistic evaluation settings for these workflows. HealthAdminBench~\cite{bedi2026healthadminbench} evaluates agents on healthcare administrative tasks spanning EHR systems, insurance portals, and fax interfaces, while MedCUA-Bench~\cite{yu2026medcua} evaluates clinical workflows in environments reconstructed from product manuals and open-source medical systems. Although these benchmarks advance the evaluation of computer-use agents in healthcare, their predefined environments and tasks cannot fully represent the diversity of workflows, patient contexts, and interface configurations across healthcare institutions. Scalable evaluation therefore requires methods for constructing additional environments and tasks from available evidence of real-world workflows.

This need is particularly important in healthcare, where proprietary software and sensitive patient information restrict access to production applications. The healthcare software market is also highly concentrated, with three developers supplying EHR systems to more than 80\% of U.S. non-federal acute care hospitals~\cite{barker2026ehr}. Reconstructing interactive environments from recordings of widely deployed applications could enable reproducible evaluation without requiring researchers to access production systems. Building on the motivation for realistic healthcare computer-use benchmarks established by HealthAdminBench, \PluginName{} addresses the complementary challenge of constructing new benchmarks from screen recordings alone while systematically assessing environment fidelity and task quality.

\section{\PluginName{}}\label{sec:method}

\subsection{Problem Formulation}\label{objective}
Given a set of screen recordings $R = \{r_1, \ldots, r_k\}$ of an application, the objective is to construct a reproducible benchmark comprising an interactive environment $E$, a set of evaluation tasks $\mathcal{T}$, and a corresponding set of task verifiers $\mathcal{V}$ that assess whether agents complete the tasks correctly. Specifically, we seek a benchmark construction method $\mathcal{M}$ such that
\[
\mathcal{M}(R) = (E, \mathcal{T}, \mathcal{V}).
\]

We model the environment as a state transition function $E: \mathcal{S} \times \mathcal{A} \rightarrow \mathcal{S}$, where $\mathcal{S}$ is the set of environment states and $\mathcal{A}$ is the set of available interface actions. Given a state $s_t \in \mathcal{S}$ and an action $a_t \in \mathcal{A}$, the environment transitions to $s_{t+1} = E(s_t, a_t)$.

The set of evaluation tasks is defined as
\[
\mathcal{T} = \{(G_i, \pi_i^*)\}_{i=1}^{m},
\]
where each task $i$ consists of a goal $G_i$ specifying the work to be performed and a golden path $\pi_i^*$ specifying an executable sequence of interface actions that achieves the intended outcome.

The corresponding set of task verifiers is defined as
\[
\mathcal{V} = \{V_i\}_{i=1}^{m},
\]
where each verifier $V_i$ assesses whether an agent has successfully completed task $i$. Each verifier comprises $n_i$ deterministic checks on the final environment state and, where applicable, judging rubrics for free-text responses. Each deterministic check $f_{i,j}$ evaluates a query $q_{i,j}$ against an expected value $v_{i,j}$:
\[
f_{i,j}(s) =
\begin{cases}
1, & \text{if } q_{i,j}(s) = v_{i,j},\\
0, & \text{otherwise.}
\end{cases}
\]
The objective is to construct an environment that faithfully reproduces the recorded interfaces and interactions, together with evaluation tasks whose verifiers accept correct executions and reject incorrect outcomes. Because recordings capture only observed behavior, the objective is not to recover the complete underlying application, but to reconstruct the functionality necessary to support reproducible evaluation of the recorded workflows.

\subsection{System Overview}

We introduce \PluginName{}, a multi-agent orchestrator implemented as a Claude Code plugin~\citep{anthropic2025claudecode} that addresses the benchmark construction problem defined above. Given screen recordings of an application, \PluginName{} reconstructs an interactive environment and generates evaluation tasks with executable solutions and verifiable outcomes, without requiring access to the original application, its source code, or its accessibility tree.

\begin{figure*}[t]
    \centering
    \includegraphics[width=\linewidth]{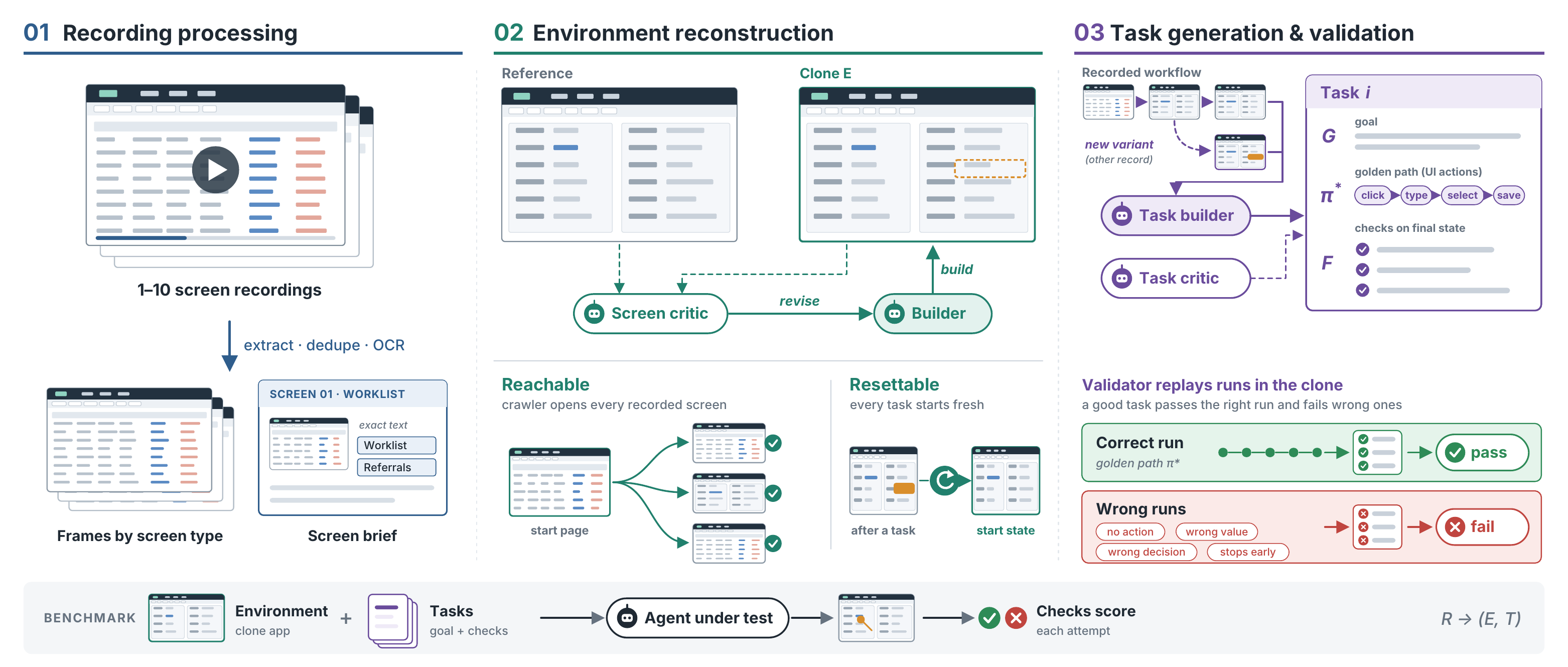}
    \caption{Overview of \PluginName{}. Screen recordings are processed to reconstruct interactive environments and generate verifiable evaluation tasks. Independent agents review reconstructed interfaces and generated tasks, while deterministic scripts validate task execution and environment reproducibility.}
    \label{fig:process}
\end{figure*}

\begin{figure*}[t]
    \centering
    \includegraphics[width=\linewidth]{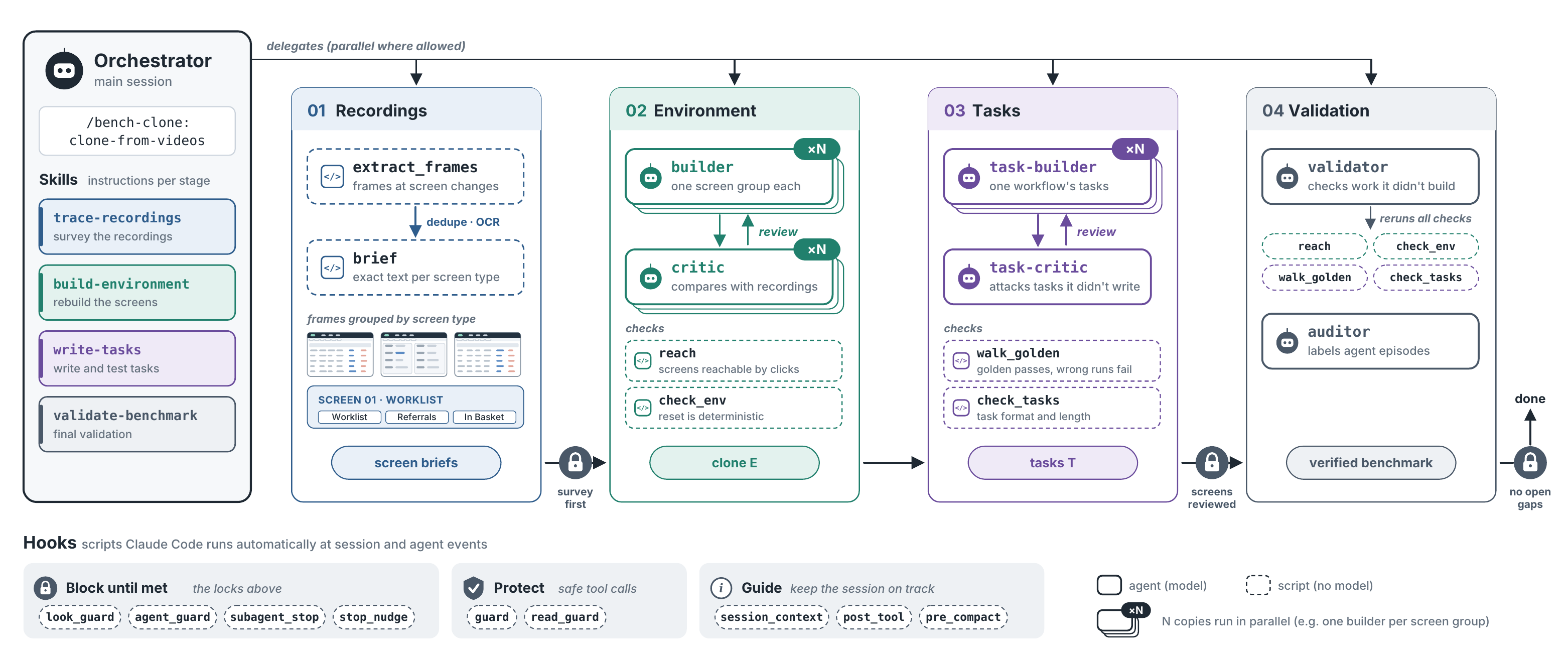}
    \caption{Multi-agent architecture of \PluginName{}. An orchestrator coordinates six specialized agents for environment reconstruction, interface review, task generation, task review, task validation, and scoring analysis. Skills provide stage-specific instructions, execution hooks coordinate dependencies and validation requirements, and deterministic scripts perform reproducible checks.}
    \label{fig:multiagent}
\end{figure*}

\PluginName{} operates in three stages (\Cref{fig:process}): (1) recording processing, (2) environment reconstruction, and (3) task generation and validation. An orchestrator coordinates specialized agents responsible for constructing environments, reviewing reconstructed interfaces, generating evaluation tasks, and validating task outcomes (\Cref{fig:multiagent}). To reduce reliance on agent self-assessment, \PluginName{} separates construction from validation: independent critic agents compare reconstructed interfaces against the recordings, while task-critic and validation agents assess generated tasks and test whether their respective task verifiers distinguish correct from incorrect outcomes. Specialized skills, execution hooks, and deterministic scripts coordinate these processes and enforce validation requirements.

\paragraph{Multi-agent architecture}
\PluginName{} employs an orchestrator that delegates construction and validation to six specialized agents (\Cref{fig:multiagent}):

\begin{itemize}
    \item \textbf{Builder:} Reconstructs application interfaces and iteratively revises them based on critic feedback.
    \item \textbf{Critic:} Independently compares reconstructed interfaces against the original recordings to identify discrepancies in content, layout, and styling.
    \item \textbf{Task-builder:} Generates evaluation tasks from recorded workflows and executes their task checks.
    \item \textbf{Task-critic:} Independently reviews generated tasks for inconsistencies, incomplete validation, and goals that reveal the expected outcomes.
     \item \textbf{Validator:} Independently tests task verifiers against correct and incorrect executions to identify cases where erroneous outcomes receive credit.
    \item \textbf{Auditor:} Analyzes agent execution episodes and their resulting task scores.
\end{itemize}

This separation of responsibilities enables independent review of reconstructed environments and generated tasks, reducing reliance on the agents that produced them to assess their own outputs.

\paragraph{Skills and workflow control}
Five skills provide stage-specific instructions for benchmark construction. The \texttt{clone-from-videos} skill serves as the entry point and coordinates the overall workflow, while \texttt{trace-recordings}, \texttt{build-environment}, \texttt{write-tasks}, and \texttt{validate-benchmark} guide recording analysis, environment reconstruction, task generation, and validation, respectively.

Execution hooks~\citep{anthropic2025hooks} enforce dependencies between stages and prevent the orchestrator from bypassing required review steps. In unattended builds, hooks restrict interface implementation until all recordings have been surveyed, require builder agents to be delegated before the orchestrator writes interface code, and delay full-environment testing until all recorded screen types have been constructed and reviewed by critic agents. They also block completion when validation gaps remain. Builder, critic, and validator agents execute in parallel where permitted, while hooks provide stage-specific guidance and restore instructions after context compaction. To avoid indefinite blocking, restrictions can be released after predefined refusal counts or time and resource limits.

\paragraph{Automated validation and completion}
Deterministic scripts complement agent-based review by performing operations such as frame extraction, browser navigation, golden-path replay, and environment reset verification. Users invoke \PluginName{} through the \texttt{/bench-clone:clone-from-videos} command, providing screen recordings and a brief request. The orchestrator then coordinates the three construction stages, allowing task generation and interface review to overlap where appropriate.

The system evaluates completion against four criteria: (1) all recorded screen types are implemented and reachable; (2) independent interface review reports no outstanding discrepancies; (3) each recorded workflow has at least one evaluation task that passes validation; and (4) resetting the environment restores a consistent initial state. These criteria support reproducible evaluation, although bounded execution limits may terminate a build before all requirements are satisfied. Even when all criteria are met, they do not guarantee complete fidelity to unobserved application behavior or exhaustive detection of incorrect task executions.

\subsection{Recording Processing}

\PluginName{} accepts one to ten screen recordings of an application and processes them to identify the interfaces and interactions needed to reproduce the recorded workflows. A preparation script extracts frames at screen transitions, removes near-duplicate frames, and groups the remaining frames by screen type. It then uses optical character recognition (OCR) to extract interface text and generates a structured brief for each screen type, containing references to the original frames and the exact text to preserve.

These briefs provide builder agents with both visual references and explicit content requirements, reducing reliance on visual interpretation alone. During environment reconstruction, builder agents use the full-resolution frames and corresponding briefs to reproduce the recorded interfaces, with instructions to preserve interface text rather than omit, invent, or reword it.

\subsection{Environment Reconstruction}

Using the extracted frames and screen briefs, the orchestrator delegates environment reconstruction to builder agents. These agents implement the recorded interfaces, navigation structure, and interactions required to reproduce the observed workflows.

\paragraph{Interface fidelity.}
To improve reconstruction fidelity, \PluginName{} separates environment construction from review. Independent critic agents compare screenshots of reconstructed interfaces against corresponding frames from the original recordings, identifying discrepancies in text, layout, and styling. Builder agents iteratively revise the environment based on this feedback, while the system tracks unresolved discrepancies across review iterations. This separation reduces reliance on agent self-evaluation, which prior work has shown can be unreliable without external feedback~\citep{huang2024selfcorrect,kamoi2024selfcorrect}.

\paragraph{Navigation and reproducibility.}
Visual fidelity alone does not establish whether the reconstructed environment supports the recorded workflows. An automated crawler, inspired by Crawljax~\citep{mesbah2012crawljax}, navigates the reconstructed environment to verify that each recorded screen type is reachable from the initial state. The system also checks that resetting the environment restores a consistent initial state, enabling evaluation tasks to be executed independently and reproducibly.

\subsection{Task Generation and Validation}

\begin{figure*}[t]
    \centering
    \includegraphics[width=\linewidth]{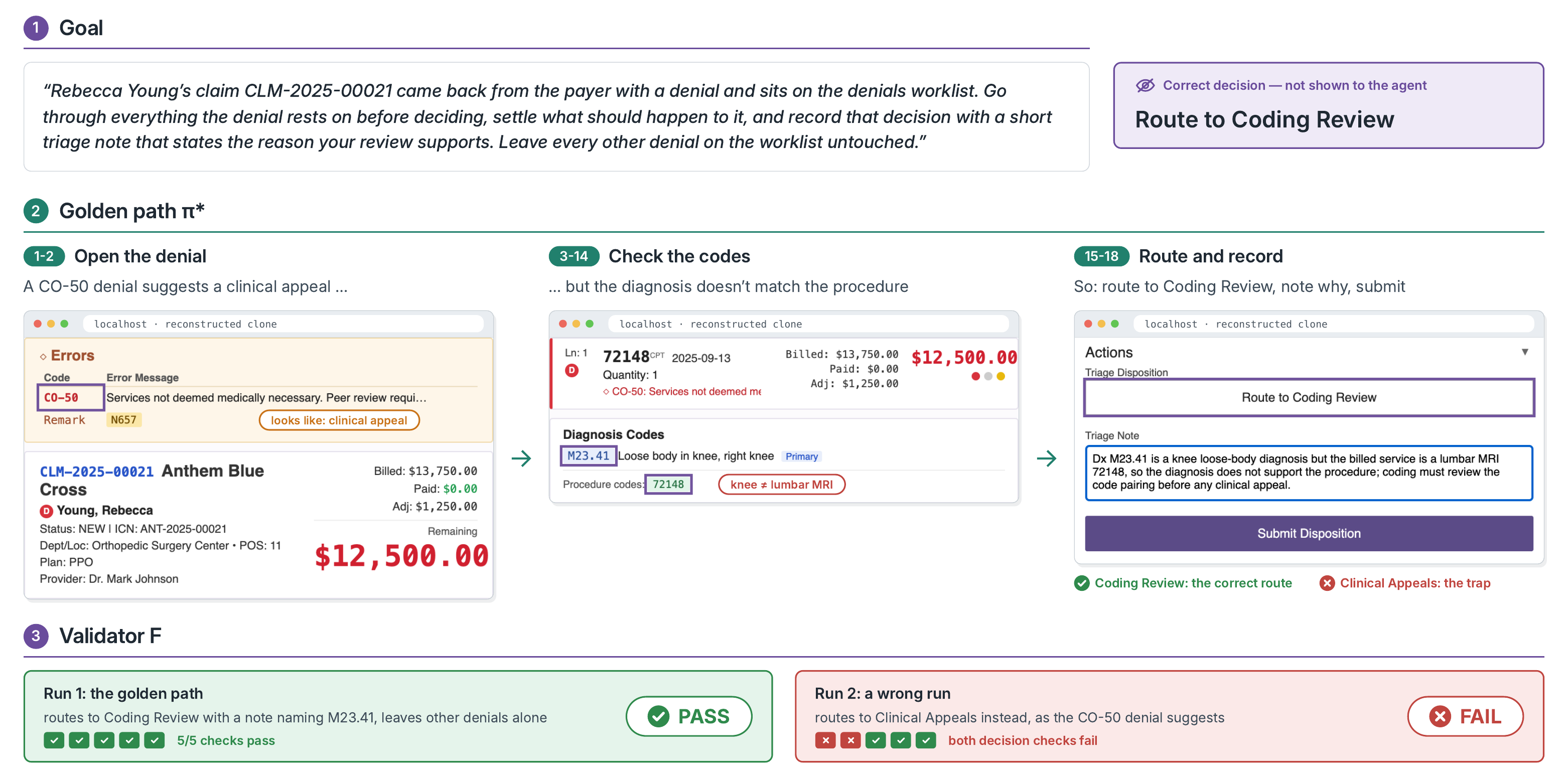}
    \caption{Example evaluation task generated by \PluginName{} with Claude Haiku 5.5 for appeals and denials. The task requires an agent to determine how to route a denied claim without revealing the expected decision. Although the denial code suggests a clinical appeal, a mismatch between the billed procedure and diagnosis requires coding review. The figure shows the goal and the screens of the reconstructed environment along the golden path. Its validator has five state-based checks and two rubrics; the golden path satisfies all state checks, while the tested incorrect outcomes fail at least one check.}
    \label{fig:task}
\end{figure*}

For each recorded workflow, \PluginName{} generates an evaluation task that reproduces the workflow using the original recorded data, along with up to two additional tasks using different records that lead to the same outcome. The system also generates novel tasks that reuse the recorded interfaces for alternative workflows, such as correcting inconsistencies across records or handling exception cases. For tasks requiring a decision, the goal does not disclose the expected outcome, requiring agents to infer the appropriate action from the available information.~\Cref{fig:task} illustrates a generated task with its golden path, and Appendix~\Cref{app:tasks} gives a complete task with its validation checks.

Following the problem formulation, each evaluation task $i$ consists of a goal $G_i$ and a golden path $\pi_i^*$, together with a corresponding verifier $V_i$ that assesses whether an agent has completed the task correctly. Each verifier comprises deterministic checks on the final environment state and, where applicable, judging rubrics for free-text responses. For construction-time validation, we define the deterministic component of the task verifier as
\[
V_i(s) = \prod_{j=1}^{n_i} f_{i,j}(s),
\]
where $V_i(s)=1$ if all deterministic checks pass and $V_i(s)=0$ otherwise. Free-text responses are evaluated separately using judging rubrics and are not included in $V_i$. The generated task verifier $V_i$ is retained as part of the benchmark and used to grade task outcomes during downstream agent evaluation.

\textbf{Golden-path testing.}
To verify that each generated task can be completed, \PluginName{} resets the environment and replays its golden path through the reconstructed interface. Let $s_i^+$ denote the final state reached by executing $\pi_i^*$. The task verifier must accept this execution:

\[
V_i(s_i^+) = 1.
\]

This establishes that the task is executable and that its verifier accepts the intended outcome.

\textbf{Mutation-based testing.}
A successful golden path does not establish whether a verifier can distinguish correct from incorrect outcomes. To assess verifier reliability, \PluginName{} constructs a set of negative test states $\mathcal{N}_i$, including unchanged states, states with incorrectly entered values, and states with reversed decisions where applicable. It also tests incomplete executions, such as navigating through the interface without completing the required actions.

Inspired by mutation testing~\citep{demillo1978hints,jia2011mutation} and best practices for agentic benchmarks~\citep{zhu2025abc}, the system replays these incorrect executions or constructs their final states directly. Each negative test must be rejected by the verifier:

\[
V_i(s^-) = 0,
\qquad \forall s^- \in \mathcal{N}_i.
\]

Together, positive and negative validations assess whether generated tasks can be completed successfully and whether their verifiers detect the incorrect outcomes. These validations do not guarantee detection of every possible execution error.

\textbf{Task consistency.}
Besides validation, \PluginName{} enforces three consistency requirements. First, all values required to complete a task must be available in its goal, initial data, or interface before they are entered. Second, task verifiers must check all stored values entered during the golden path, with free-text responses evaluated separately using rubrics. Third, completing one task must not cause another task to receive credit. These requirements reduce incorrect task scores and improve the reliability of generated evaluations.

\section{Evaluation Methodology}\label{sec:exp}

\subsection{Experimental Setup}

We evaluate \PluginName{} on all three healthcare applications from HealthAdminBench: prior authorization, appeals and denials, and orders for durable medical equipment (DME). These applications represent complex administrative workflows involving patient worklists, electronic health record (EHR) views, insurance portals, and fax interfaces, populated with synthetic data.

For each application and each of three base models (Claude Opus 5.5, Claude Sonnet 5.5, and Claude Haiku 5.5), we generate two builds using Claude Code with \PluginName{} and two baseline builds using Claude Code alone, yielding twelve builds per base model. All builds use the same effort level (medium), Claude Code version (2.1.330), screen recordings, and construction request. The recordings consist of an application tour and one or two complete workflows. The construction request (Appendix~\Cref{app:request}) asks for a resettable interactive environment and evaluation tasks derived from the recordings, each containing a goal, step-by-step instructions, a golden path, and a verifier with at least five checks or rubrics. It further specifies that a typical incorrect outcome should receive a score below 60\%.

All builds operate under identical time and resource limits. Each \PluginName{} build stopped before confirming all four internal completion criteria described in~\Cref{sec:method} and was evaluated in its resulting state. This setup enables a controlled comparison of benchmark construction with and without the plugin while holding the base model, input recordings, instructions, and computational constraints fixed.

\subsection{Evaluation Framework}\label{sec:eval}

We assess the resulting benchmarks along two dimensions: \textbf{environment fidelity}, which measures how faithfully reconstructed environments reproduce recorded interfaces and interactions, and \textbf{task quality}, which assesses the coverage, executability, and reliability of generated evaluation tasks. We use four measures of environment fidelity and three measures of task quality, including two newly proposed measures: \textbf{task preference} and \textbf{check strength}.~\Cref{tab:design} connects the challenges addressed by \PluginName{} to its corresponding mechanisms and evaluation measures.

\begin{table*}[t]
\centering
\small
\renewcommand{\arraystretch}{1.3}
\begin{tabularx}{\textwidth}{@{\hspace{4pt}}
  >{\raggedright\arraybackslash}p{0.2\textwidth}@{\colgap}
  >{\raggedright\arraybackslash}X@{\colgap}
  >{\raggedright\arraybackslash}p{0.3\textwidth}@{}}
\toprule
\textbf{Challenge} & \textbf{\PluginName{} mechanism} & \textbf{Measure} \\
\midrule
\grouprow{env}{Environment fidelity}{Does the clone match the recordings?}
Missing or altered text & Builders get each screen's OCR-extracted text & Text match \\
Visual mismatch & A critic agent compares screens to the recordings & Visual preference, visual similarity \\
Unreachable screens & A crawler clicks through the clone to every screen & Reach \\[3pt]
\grouprow{task}{Task quality}{Are tasks solvable, and do checks catch wrong outcomes?}
Incomplete coverage & One task per recorded workflow, critic-reviewed & Task preference$^{\dagger}$ \\
Non-executable tasks & Golden paths are replayed in a freshly reset clone & Working tasks \\
Wrong outcomes pass & Deliberately wrong runs must fail the checks & Check strength$^{\dagger}$, working tasks \\
\bottomrule
\end{tabularx}
\caption{Benchmark-construction challenges, the \PluginName{} mechanism that targets each, and the measure that evaluates it. $^{\dagger}$New measures introduced in this work.}
\label{tab:design}
\end{table*}

\paragraph{Reference screens and matching.}
We select reference screens from the recordings of each application, retaining those for which a crawler identifies a closely matching screen in the original application. Because many screens differ only in record-specific content, we group similar screens into clusters and assign equal weight to each cluster to prevent near-identical screens from disproportionately influencing the evaluation.

For each reconstructed environment, an automated crawler navigates the interface, captures reachable screens, and matches them to reference screens using headings, tabs, and buttons. In visual preference comparisons, a build loses a reference screen when only the other build contains a matching screen; when neither build contains a match, the screen counts as a tie. Missing reference screens receive zero scores for visual similarity and text match.

\paragraph{Environment fidelity.}
We evaluate environment fidelity using four complementary measures:

\begin{itemize}

    \item \textbf{Visual preference.} An LLM judge compares a reference screen from the original recording against corresponding screenshots from a \PluginName{} build and a baseline build. Without knowing which method produced either reconstruction, the judge selects the screenshot that more closely matches the reference. To mitigate position bias~\citep{zheng2023judging,shi2025position}, we repeat the comparison three times in each presentation order. A build wins a screen only when a majority of judgments in both orders favors it; otherwise, the screen is treated as a tie.

    \item \textbf{Visual similarity.} We measure pixel-level fidelity by averaging three components: overlap between content regions, color similarity based on CIEDE2000~\citep{sharma2005ciede2000}, and structural similarity across image scales using multi-scale SSIM~\citep{wang2004ssim,wang2003msssim}. The measure follows the principles of screenshot-based evaluation used in Design2Code~\citep{si2025design2code}.

    \item \textbf{Text match.} We assess whether reconstructed interfaces preserve recorded text and its approximate position. Optical character recognition (OCR) extracts text lines from reference and reconstructed screens, which are paired using a one-to-one matching procedure inspired by Design2Code~\citep{si2025design2code}. Each matched pair is weighted by textual similarity, measured using normalized edit distance~\citep{levenshtein1966binary}, and spatial proximity. The final score is a weighted F1 measure that penalizes both missing and additional interface text.

    \item \textbf{Reach.} We measure the number of reference screens reachable from the reconstructed environment's initial page. Inspired by reachability testing in Crawljax~\citep{mesbah2012crawljax}, a screen is considered reached when the reference and reconstructed screens are mutual best matches with a matching score of at least 0.5.

\end{itemize}

\paragraph{Task quality.}
We evaluate task quality using three complementary measures:

\begin{itemize}

    \item \textbf{Task preference.} An LLM judge compares the original HealthAdminBench task for a recorded workflow against the complete task sets generated by \PluginName{} and the baseline. The judge selects the better task set based on coverage of the recorded workflow, whether generated goals preserve decisions for the agent to make, and whether verifiers and rubrics assess the required changes. To mitigate position bias, we repeat each comparison three times in both presentation orders. A workflow counts as a tie unless a majority of judgments in both orders favors the same build. We evaluate two recorded workflows for prior authorization and one for each remaining application.

    \item \textbf{Working tasks.} An independent scoring script counts generated tasks that can be executed and correctly scored in the reconstructed environment. A task must reset to an unsolved initial state, accept its golden path and correct variations, reject simulated incorrect executions (Appendix~\Cref{app:working}), and provide all required input values through the task goal, initial data, or interface. Neither construction method has access to the scoring script. Although \PluginName{} performs similar checks internally, this measure independently evaluates the resulting tasks.

    \item \textbf{Check strength.} We measure the proportion of stored values entered or selected during a task's golden path that are verified by deterministic checks. A value is considered verified when entering it incorrectly causes the corresponding check to fail. Free-text responses, such as generated notes, are excluded because they are evaluated separately using judging rubrics. Conceptually related to mutation adequacy in software testing \citep{jia2011mutation}, check strength quantifies the coverage of deterministic task verifiers by measuring whether deliberately incorrect values are detected during task evaluation.

\end{itemize}

Both preference measures use Claude Opus 5.5 as the judge (Appendix~\Cref{app:judges}). The judge is not informed which method generated each reconstruction or task set.

\subsection{Comparison Protocol}

For each application and base model, we compare each of the two \PluginName{} builds with each of the two baseline builds, yielding four build pairs per application and twelve pairs per base model. We determine a win, tie, or loss for \PluginName{} on each of six evaluation measures. Reach is reported separately and excluded from these comparisons.

For \textbf{visual preference} and \textbf{task preference}, we calculate the proportion of reference screens or recorded workflows, respectively, for which the LLM judge prefers \PluginName{} over the baseline. Each screen or workflow receives a score of 1 if the judge prefers \PluginName{}, 0 if the judge prefers the baseline, and 0.5 if neither is preferred. We average these scores across screens or workflows, assigning equal weight to each screen cluster for visual preference. A preference share of at least 0.65 constitutes a win for \PluginName{}, a share of at most 0.35 constitutes a loss, and intermediate values constitute a tie.

For \textbf{visual similarity} and \textbf{text match}, the build with the higher score wins, unless the scores differ by no more than 5\% of the larger score, in which case the comparison is a tie. For \textbf{working tasks}, the build with more working tasks wins, while for \textbf{check strength}, the build with the higher proportion of verified values wins.

We also apply predefined rules for builds with insufficient tasks or values to evaluate. A \PluginName{} build with no tasks loses on task preference, working tasks, and check strength, while a baseline build with no tasks loses these comparisons only when \PluginName{} generates at least one working task. A build that generates tasks but no applicable entered values loses on check strength. Finally, wins on working tasks and check strength must remain valid under sensitivity checks that respectively relax one task check and include free-text values in the check-strength denominator; otherwise, the comparisons are treated as ties.

These rules yield twelve comparisons per measure and 72 metric-level comparisons per base model. Because the same builds contribute to multiple pairwise comparisons, these outcomes summarize relative performance rather than representing 72 independent experimental runs.

\section{Results}\label{sec:results}

\subsection{Overall Performance}

Under matched experimental conditions, \PluginName{} outperforms Claude Code alone across most evaluation measures. With Claude Opus 5.5, across twelve paired build comparisons and six measures, \PluginName{} wins 71 of 72 metric-level comparisons, ties 1, and loses none (\Cref{tab:pairs}). It wins 67 of 72 with Claude Sonnet 5.5 and 64 of 72 with Claude Haiku 5.5. Improvements are particularly consistent for \textbf{environment fidelity}, with 36 wins across 36 comparisons for Claude Opus 5.5 and Claude Sonnet 5.5, and 36 for Claude Haiku 5.5. For \textbf{task quality}, \PluginName{} achieves 35 wins, 1 tie, and no losses across 36 comparisons with Claude Opus 5.5. It achieves 31 wins with Claude Sonnet 5.5 and 28 with Claude Haiku 5.5.

\begin{table}[t]
\centering
\small
\renewcommand{\arraystretch}{1.25}
\begin{tabular*}{\textwidth}{@{}l@{\extracolsep{\fill}}ccccccc@{}}
\toprule
& \multicolumn{3}{c}{\textcolor{envc}{\textbf{Environment fidelity}}}
& \multicolumn{3}{c}{\textcolor{taskc}{\textbf{Task quality}}} & \\
\cmidrule(lr){2-4}\cmidrule(lr){5-7}
& \shortstack{Visual\\preference} & \shortstack{Visual\\similarity} & \shortstack{Text\\match}
& \shortstack{Task\\preference} & \shortstack{Working\\tasks} & \shortstack{Check\\strength}
& \textbf{Overall} \\
\midrule
\multicolumn{8}{@{}l}{\textit{Claude Opus 5.5}} \\
\textbf{Wins} (of 12) & \textbf{12} & \textbf{12} & \textbf{12} & \textbf{11} & \textbf{12} & \textbf{12} & \textbf{71}/72 \\
\textcolor{black!55}{Ties / losses} & \textcolor{black!55}{0 / 0} & \textcolor{black!55}{0 / 0} & \textcolor{black!55}{0 / 0} & \textcolor{black!55}{1 / 0} & \textcolor{black!55}{0 / 0} & \textcolor{black!55}{0 / 0} & \textcolor{black!55}{1 / 0} \\
\midrule
\multicolumn{8}{@{}l}{\textit{Claude Sonnet 5.5}} \\
\textbf{Wins} (of 12) & \textbf{12} & \textbf{12} & \textbf{12} & \textbf{9} & \textbf{12} & \textbf{10} & \textbf{67}/72 \\
\textcolor{black!55}{Ties / losses} & \textcolor{black!55}{0 / 0} & \textcolor{black!55}{0 / 0} & \textcolor{black!55}{0 / 0} & \textcolor{black!55}{1 / 2} & \textcolor{black!55}{0 / 0} & \textcolor{black!55}{2 / 0} & \textcolor{black!55}{3 / 2} \\
\midrule
\multicolumn{8}{@{}l}{\textit{Claude Haiku 5.5}} \\
\textbf{Wins} (of 12) & \textbf{12} & \textbf{12} & \textbf{12} & \textbf{12} & \textbf{10} & \textbf{6} & \textbf{64}/72 \\
\textcolor{black!55}{Ties / losses} & \textcolor{black!55}{0 / 0} & \textcolor{black!55}{0 / 0} & \textcolor{black!55}{0 / 0} & \textcolor{black!55}{0 / 0} & \textcolor{black!55}{1 / 1} & \textcolor{black!55}{6 / 0} & \textcolor{black!55}{7 / 1} \\
\bottomrule
\end{tabular*}
\caption{Wins of \PluginName{} over Claude Code alone, with ties and losses below. Each measure compares twelve matched build pairs (three applications $\times$ two plugin builds $\times$ two baseline builds), giving 72 comparisons per base model. Reach is reported separately and excluded.}
\label{tab:pairs}
\end{table}

\Cref{tab:headline} summarizes performance with Claude Opus 5.5 across the three healthcare applications, aggregating the two builds of each method. Visual similarity and text match are averaged across builds, while reach, working tasks, and check strength report pooled counts. These application-level summaries complement the pairwise comparisons used to determine wins and losses (Appendix~\Cref{app:perbuild}).

\begin{table*}[t]
\centering
\small
\renewcommand{\arraystretch}{1.3}
\begin{tabular*}{\textwidth}{@{\hspace{4pt}}l@{\extracolsep{\fill}}cccccc@{\hspace{4pt}}}
\toprule
& \multicolumn{2}{c}{\textbf{Prior authorization}}
& \multicolumn{2}{c}{\textbf{Appeals and denials}}
& \multicolumn{2}{c}{\textbf{DME}} \\
\cmidrule(lr){2-3}\cmidrule(lr){4-5}\cmidrule(lr){6-7}
\textbf{Measure}
& \textbf{Ours} & \dn{Plain} & \textbf{Ours} & \dn{Plain} & \textbf{Ours} & \dn{Plain} \\
\midrule
\bandrow{env}{Environment fidelity}{7}
Visual similarity \dn{(0--1)}   & \textbf{0.73} & 0.45 & \textbf{0.89} & 0.65 & \textbf{0.85} & 0.63 \\
Text match \dn{(0--1)}          & \textbf{0.75} & 0.34 & \textbf{0.88} & 0.59 & \textbf{0.78} & 0.64 \\
Reach \dn{(screens reached)}    & \textbf{13}\dn{/16} & 11\dn{/16} & \textbf{4}\dn{/4} & 3\dn{/4} & \textbf{10}\dn{/10} & 9\dn{/10} \\[2pt]
\bandrow{task}{Task quality}{7}
Working tasks \dn{(of generated)} & \textbf{13}\dn{/15} & 2\dn{/14} & \textbf{12}\dn{/13} & 4\dn{/19} & \textbf{12}\dn{/12} & 0\dn{/13} \\
Check strength \dn{(\% verified)} & \textbf{100\%} & 82\% & \textbf{100\%} & 77\% & \textbf{100\%} & 63\% \\
\bottomrule
\end{tabular*}
\caption{Results per application with Claude Opus 5.5, pooling the two builds of \PluginName{} (Ours) and of Claude Code alone (Plain). Visual similarity and text match are averaged over builds; reach and working tasks are pooled counts; check strength is the share of entered values that a check verifies. Bold marks the better method. Visual and task preference are pairwise and appear in~\Cref{tab:pairs}.}
\label{tab:headline}
\end{table*}

\subsection{Environment Fidelity}

\PluginName{} consistently produces environments that more closely reproduce the original recordings. With Claude Opus 5.5 and Claude Sonnet 5.5, it wins all twelve comparisons on visual preference, visual similarity, and text match, and with Claude Haiku 5.5, it wins 36 of these 36 comparisons. In eight of the twelve visual preference comparisons with Claude Opus 5.5, at least 97\% of the weighted reference screens favor \PluginName{}. This holds in all twelve with Claude Sonnet 5.5 and in eight with Claude Haiku 5.5.

The improvements are evident across all three applications with every base model (\Cref{tab:headline} and Appendix~\Cref{app:perbuild}). For prior authorization with Claude Opus 5.5, mean text match increases from 0.34 to 0.75, while for appeals and denials it increases from 0.59 to 0.88. With Claude Sonnet 5.5, mean text match increases from 0.31 to 0.90 for prior authorization and from 0.26 to 0.82 for appeals and denials. With Claude Haiku 5.5, it increases from 0.31 to 0.65 and from 0.20 to 0.65. Visual similarity also improves across all three applications, and with Claude Opus 5.5 the largest absolute increase is observed for prior authorization (0.45 to 0.73). These findings indicate that \PluginName{} better preserves the visual structure and textual content of recorded interfaces.

The reconstructed environments also support broader navigation coverage. For DME with Claude Opus 5.5, the two \PluginName{} builds collectively reach all 10 reference-screen instances, compared with 9 of 10 for Claude Code alone. Similarly, for appeals and denials, \PluginName{} reaches all 4 reference-screen instances across its two builds, compared with three for the baseline. With Claude Opus 5.5 and Claude Sonnet 5.5, the \PluginName{} builds reach more reference-screen instances than the baseline builds in all three applications.

\begin{figure*}[t]
\centering

\newcommand{\bigscreen}[1]{%
\IfFileExists{figures/fig3_#1.png}{%
\includegraphics[
width=0.49\linewidth,
height=0.2756\linewidth,
keepaspectratio
]{figures/fig3_#1.png}%
}{%
\fbox{\parbox[c][0.27\linewidth][c]
{0.46\linewidth}
{\centering\scriptsize\detokenize{figures/fig3_#1.png}}}%
}%
}

\setlength{\tabcolsep}{2pt}
\renewcommand{\arraystretch}{1.2}

\begin{tabular}{@{}cc@{}}
\multicolumn{2}{c}{\small Original application} \\
\multicolumn{2}{c}{\bigscreen{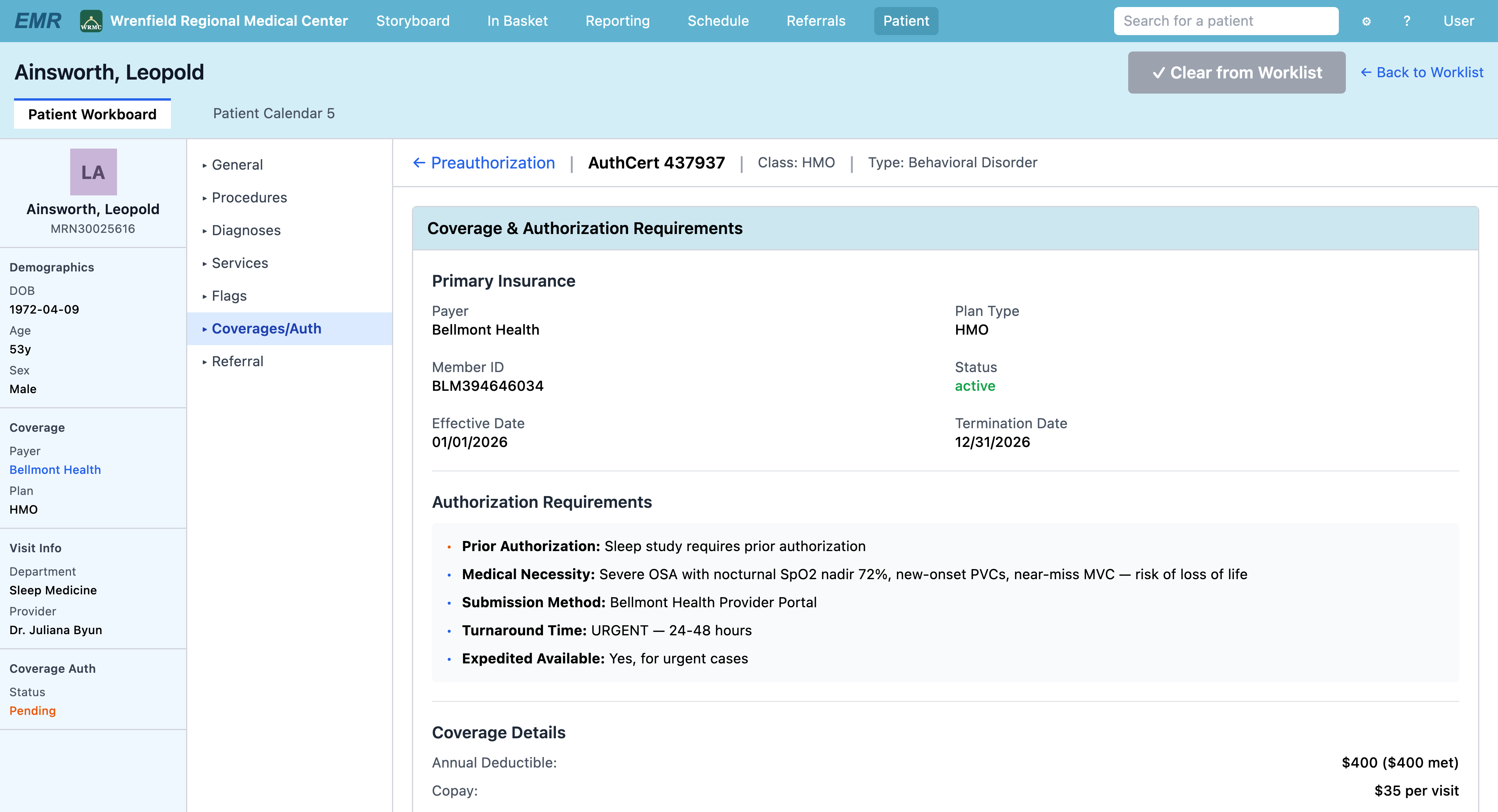}} \\
\noalign{\vskip 4pt}
\small Plain Claude Code & \small \PluginName{} \\
\bigscreen{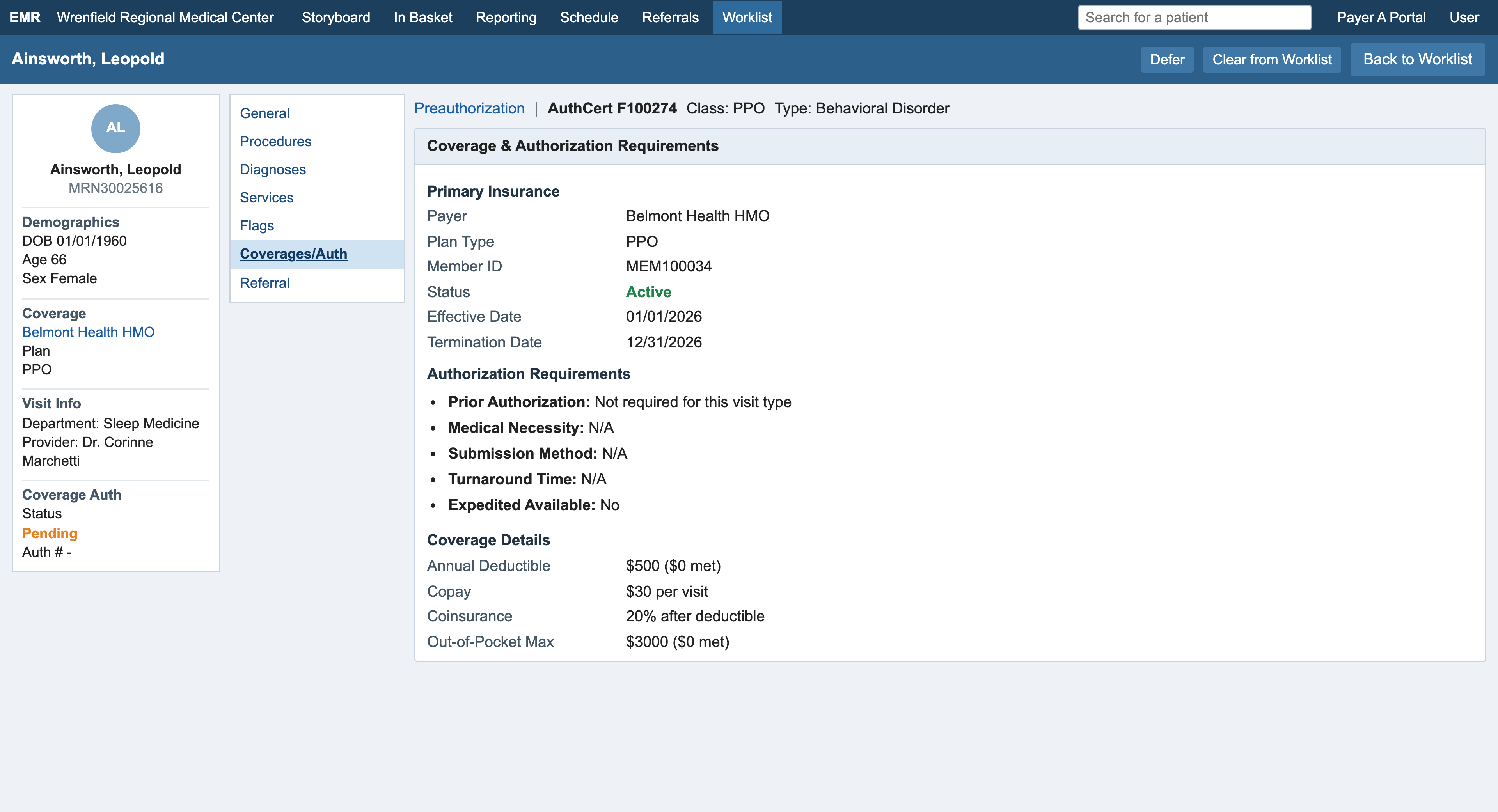} & \bigscreen{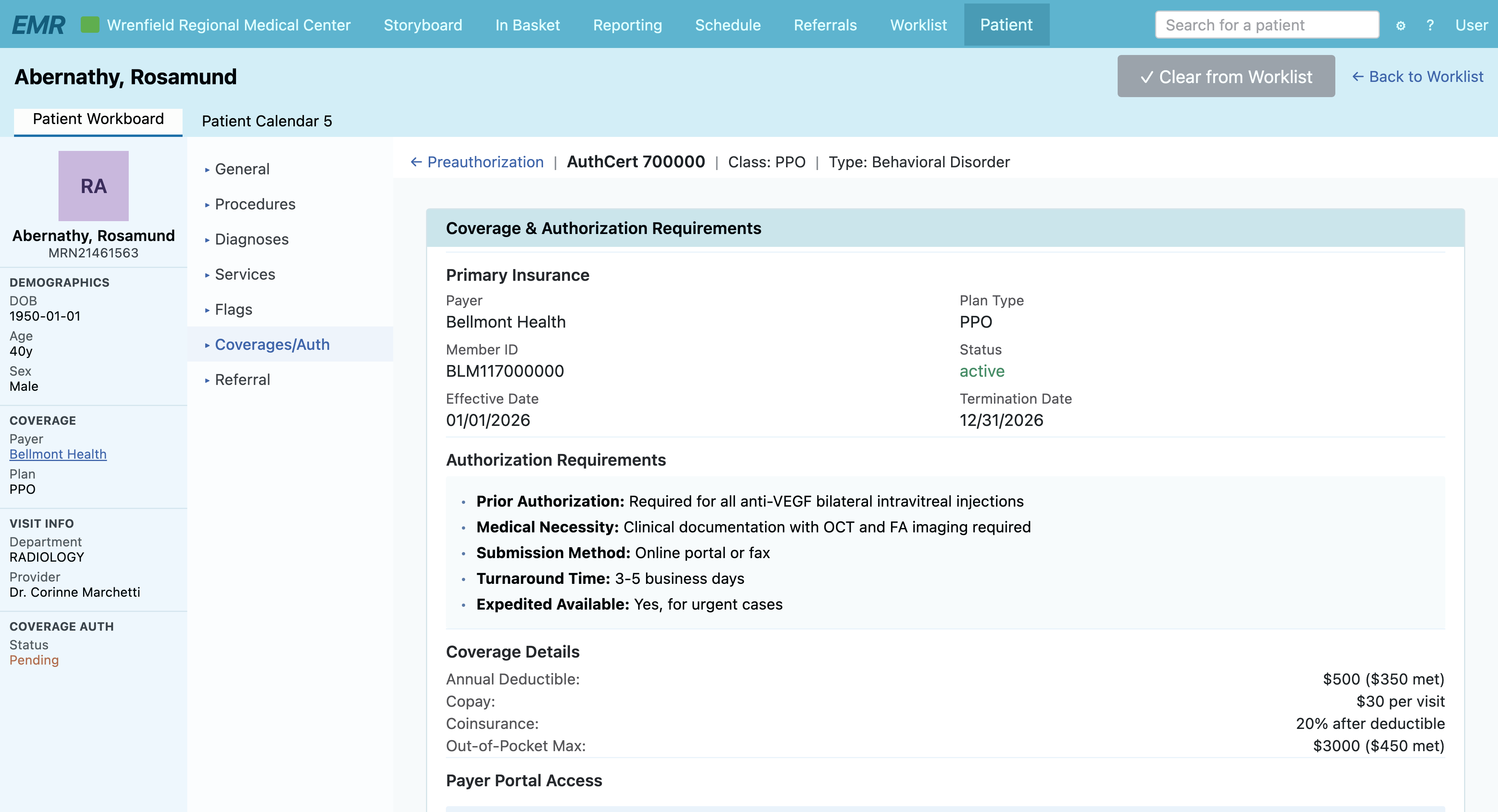} \\
\end{tabular}

\caption{Comparison of the prior authorization interface in the original application (top), a baseline reconstruction produced by Claude Code alone (bottom left), and a reconstruction produced with \PluginName{} (bottom right). Both reconstructions use Claude Haiku 5.5 and identical recordings. \PluginName{} more closely preserves the recorded layout and interface content, including the workbench navigation and insurance information panel. The LLM judge preferred the \PluginName{} reconstruction over both baseline builds for this reference screen.}
\label{fig:fidelity}
\end{figure*}

\Cref{fig:fidelity} provides a qualitative comparison of environment reconstruction. The baseline reconstruction of the prior authorization application omits the workbench tabs and two-column insurance information panel, whereas the \PluginName{} reconstruction more closely preserves the recorded layout. Additional examples from appeals and denials and DME are shown in~\Cref{fig:fidelity-more}.

\subsection{Task Quality}

With Claude Opus 5.5, \PluginName{} also improves generated task quality, winning eleven comparisons on task preference, all twelve on working tasks, and all twelve on check strength. The remaining task preference comparison is a tie, with no losses. With Claude Sonnet 5.5, it wins nine comparisons on task preference, all twelve on working tasks, and ten on check strength. With Claude Haiku 5.5, it wins all twelve comparisons on task preference, ten on working tasks, and six on check strength.

Across all \PluginName{} builds and base models, task verifiers check every applicable stored value entered during the golden paths. In contrast, with Claude Opus 5.5, all six baseline builds leave some entered values unchecked. For prior authorization, the pooled check-strength score increases from 94/114 for Claude Code alone to 80/80 with \PluginName{}. With Claude Sonnet 5.5, one baseline build for appeals and denials and one for DME leave some entered values unchecked, as do both baseline builds for prior authorization and one for DME with Claude Haiku 5.5. These results indicate that \PluginName{} produces more comprehensive task verifiers that are less likely to accept incorrect entered values.

Improvements in task executability are particularly apparent for DME, where \PluginName{} generates twelve working tasks across its two builds with Claude Opus 5.5, compared with none for Claude Code alone. For prior authorization, \PluginName{} generates thirteen working tasks across fifteen tasks, compared with two working tasks across fourteen baseline tasks. With Claude Sonnet 5.5, \PluginName{} generates five working tasks for DME and thirteen for prior authorization, compared with none and two for Claude Code alone. With Claude Haiku 5.5, it generates seven working tasks for DME and ten for prior authorization, compared with two and one. In both applications, the plugin increases the number of tasks that can be successfully executed and verified.

Although improvements are observed across most comparisons, their magnitude varies across applications and base models. With Claude Opus 5.5, \PluginName{} loses no comparisons. With Claude Sonnet 5.5, both losses occur in appeals and denials, where the judge prefers the baseline tasks to those of \PluginName{} Build 2 on task preference. In appeals and denials with Claude Haiku 5.5, one baseline build generates four working tasks, compared with three and four for the \PluginName{} builds. This accounts for the one loss and one tie on working tasks, despite all seven tasks generated by \PluginName{} across its two builds passing independent verification. Thus, while \PluginName{} improves the reliability of generated tasks, it does not consistently produce a larger absolute number of working tasks in every application.

\subsection{Analysis of System Mechanisms}

The observed improvements align with the benchmark construction mechanisms introduced in \PluginName{} (\Cref{tab:design}). For environment fidelity, screen briefs provide explicit textual requirements, while independent critic agents compare reconstructed interfaces against the original recordings and guide iterative revisions. These mechanisms target the missing or altered interface content observed in baseline reconstructions.

For task quality, \PluginName{} extends verification beyond successful golden-path execution by testing incorrect entered values, incomplete executions, and reversed decisions where applicable. It also requires deterministic checks to cover stored values entered during task execution. These mechanisms address a limitation of baseline task verifiers: a verifier may accept a correct execution while still leaving required values unchecked, allowing some incorrect outcomes to receive credit.

Taken together, our findings demonstrate that \PluginName{} improves environment fidelity and task quality under matched experimental conditions. By evaluating the multi-agent framework as an integrated system, our experiments establish the effectiveness of combining specialized agents for benchmark construction and validation. The observed improvements align with the intended roles of the critic and validation agents, motivating further investigation into their individual contributions.

\section{Limitations}

\PluginName{} provides an effective approach for converting screen recordings into reproducible computer-use benchmarks. Our experiments use simulated healthcare applications rather than production systems, consistent with our objective of assessing how faithfully \PluginName{} reconstructs the interfaces and workflows captured in screen recordings, regardless of the provenance of the original environment.

Nevertheless, screen recordings capture only observed interfaces and workflows, leaving unrecorded functionality and underlying application logic potentially incomplete. Publicly available tutorial videos may also contain visual noise and interruptions, motivating further investigation into the robustness of generating computer use benchmarks. Finally, although \PluginName{} tests task verifiers against correct and selected incorrect executions, further experiments could establish their reliability across a broader range of agent behaviors.

\section{Conclusion}

We introduce \PluginName{}, a multi-agent framework that reconstructs interactive environments and generates verifiable evaluation tasks from screen recordings, without requiring access to the original applications. By combining independent interface review with explicit validation of task verifiers, \PluginName{} addresses two central challenges in benchmark construction: environment fidelity and task quality. We also introduce two evaluation measures, task preference and check strength, inspired by mutation testing, to assess the coverage and reliability of generated tasks. Across three healthcare applications, \PluginName{} outperforms Claude Code alone in 71, 67, and 64 of 72 metric-level comparisons with Claude Opus 5.5, Claude Sonnet 5.5, and Claude Haiku 5.5, respectively, demonstrating improvements in interface fidelity and task reliability under matched experimental conditions. These findings suggest that screen recordings can serve as a practical foundation for constructing reproducible computer-use benchmarks, expanding evaluation to applications that researchers cannot directly access. More broadly, \PluginName{} offers a pathway toward scalable benchmark construction as software applications and agent capabilities continue to evolve.

\section{Ethical Considerations}

\PluginName{} enables computer-use agents to be evaluated in reconstructed software environments rather than directly interacting with the original applications. Such environments should be reconstructed and used with appropriate authorization, in compliance with applicable software licensing and copyright restrictions.

Because Claude Code transmits frames from screen recordings to the model provider during benchmark construction, recordings of healthcare applications may expose sensitive patient information. Whenever possible, recordings should therefore be obtained from demonstration environments populated with synthetic patient data. Benchmarks constructed from recordings containing real patient data should undergo human review for identifying information before distribution.

All screen recordings used in our experiments were obtained from HealthAdminBench, whose simulated healthcare applications contain exclusively synthetic patient data. Accordingly, our experimental evaluation of \PluginName{} was conducted without accessing proprietary production systems, infringing software copyrights, or exposing real patient information.
\section{Acknowledgements}
This work was supported in part by the National Library of Medicine of the National Institutes of Health under award number R01LM014795. 

\bibliographystyle{assets/plainnat}
\bibliography{references}

\begin{thebibliography}{37}
\providecommand{\natexlab}[1]{#1}
\providecommand{\url}[1]{\texttt{#1}}
\expandafter\ifx\csname urlstyle\endcsname\relax
  \providecommand{\doi}[1]{doi: #1}\else
  \providecommand{\doi}{doi: \begingroup \urlstyle{rm}\Url}\fi

\bibitem[{Anthropic}(2025{\natexlab{a}})]{anthropic2025claudecode}
{Anthropic}.
\newblock {Claude Code} overview.
\newblock \url{https://code.claude.com/docs/en/overview}, 2025{\natexlab{a}}.
\newblock Accessed 2026-11-12.

\bibitem[{Anthropic}(2025{\natexlab{b}})]{anthropic2025hooks}
{Anthropic}.
\newblock Hooks reference, {Claude Code} documentation.
\newblock \url{https://code.claude.com/docs/en/hooks}, 2025{\natexlab{b}}.
\newblock Accessed 2026-11-12.

\bibitem[Bai et~al.(2026)Bai, Taymanov, Zhang, Kumar, and Whitehead]{Bai2026CVPR}
Hao Bai, Alexey Taymanov, Tong Zhang, Aviral Kumar, and Spencer Whitehead.
\newblock {WebGym}: Scaling training environments for long-horizon visual web agents with realistic tasks.
\newblock In \emph{Proceedings of the IEEE/CVF Conference on Computer Vision and Pattern Recognition (CVPR)}, pages 12248--12258, June 2026.

\bibitem[Barker and Chang(2026)]{barker2026ehr}
W.~Barker and W.~Chang.
\newblock Non-federal acute care hospital electronic health record adoption, 2008--2024.
\newblock ONC Data Brief No.~83, Office of the National Coordinator for Health Information Technology, \url{https://healthit.gov/data/data-briefs/non-federal-acute-care-hospital-electronic-health-record-adoption-2008-2024/}, 2026.

\bibitem[Bedi et~al.(2026)Bedi, Welch, Steinberg, Wornow, Kim, Ahmed, Sterling, Purohit, Akram, Acosta, Nubla, Sharma, Pfeffer, Koyejo, and Shah]{bedi2026healthadminbench}
Suhana Bedi, Ryan Welch, Ethan Steinberg, Michael Wornow, Taeil~Matthew Kim, Haroun Ahmed, Peter Sterling, Bravim Purohit, Qurat Akram, Angelic Acosta, Esther Nubla, Pritika Sharma, Michael~A. Pfeffer, Sanmi Koyejo, and Nigam~H. Shah.
\newblock {HealthAdminBench}: Evaluating computer-use agents on healthcare administration tasks.
\newblock \emph{arXiv preprint arXiv:2604.09937}, 2026.
\newblock \url{https://arxiv.org/abs/2604.09937}.

\bibitem[Boisvert et~al.(2024)Boisvert, Thakkar, Gasse, Caccia, De~Chezelles, Cappart, Chapados, Lacoste, and Drouin]{boisvert2024workarenaplusplus}
L{\'e}o Boisvert, Megh Thakkar, Maxime Gasse, Massimo Caccia, Thibault Le~Sellier De~Chezelles, Quentin Cappart, Nicolas Chapados, Alexandre Lacoste, and Alexandre Drouin.
\newblock {WorkArena++}: Towards compositional planning and reasoning-based common knowledge work tasks.
\newblock In \emph{Advances in Neural Information Processing Systems}, volume~37, pages 5996--6051, 2024.

\bibitem[Bonacci and DuPont(2025)]{qi2026cuabench}
Francesco Bonacci and Dillon DuPont.
\newblock {Cua-Bench}: A framework for benchmarking, training data, and {RL} environments for computer-use agents.
\newblock Hugging Face Blog, \url{https://huggingface.co/blog/cua-ai/cua-bench}, 2025.
\newblock Published 2025-12-16. Accessed 2026-11-13.

\bibitem[Cao et~al.(2026)Cao, Ran, Wu, Guo, Chen, Li, Cao, Zhi, Yu, Li, Yang, and Xie]{cao2026guigenesis}
Yuan Cao, Dezhi Ran, Mengzhou Wu, Yuzhe Guo, Xin Chen, Ang Li, Gang Cao, Gong Zhi, Hao Yu, Linyi Li, Wei Yang, and Tao Xie.
\newblock {GUI-GENESIS}: Automated synthesis of efficient environments with verifiable rewards for {GUI} agent post-training.
\newblock \emph{arXiv preprint arXiv:2602.14093}, 2026.

\bibitem[Chae et~al.(2026)Chae, Park, and Ritter]{chae2026verienv}
Hyungjoo Chae, Jungsoo Park, and Alan Ritter.
\newblock Safe and scalable web agent learning via recreated websites.
\newblock arXiv preprint arXiv:2603.10505, 2026.

\bibitem[Chen et~al.(2026)Chen, Liu, Shen, Li, Huang, Fang, Zheng, Huang, Yang, Fu, Mei, Wu, Wen, Yang, Mao, Lin, Yu, Shen, Qiao, and Shi]{chen2026iwrbench}
Yang Chen, Minghao Liu, Yufan Shen, Yunwen Li, Tianyuan Huang, Xinyu Fang, Tianyu Zheng, Wenxuan Huang, Cheng Yang, Daocheng Fu, Jianbiao Mei, Rong Wu, Licheng Wen, Xuemeng Yang, Song Mao, Qunshu Lin, Zhi Yu, Yongliang Shen, Yu~Qiao, and Botian Shi.
\newblock {IWR-Bench}: Can {LVLMs} reconstruct interactive webpage from a user interaction video?
\newblock In \emph{International Conference on Learning Representations}, 2026.
\newblock \url{https://openreview.net/forum?id=1zOp2WPMdZ}.

\bibitem[DeMillo et~al.(1978)DeMillo, Lipton, and Sayward]{demillo1978hints}
Richard~A. DeMillo, Richard~J. Lipton, and Frederick~G. Sayward.
\newblock Hints on test data selection: Help for the practicing programmer.
\newblock \emph{Computer}, 11\penalty0 (4):\penalty0 34--41, 1978.
\newblock \doi{10.1109/C-M.1978.218136}.

\bibitem[Huang et~al.(2024)Huang, Chen, Mishra, Zheng, Yu, Song, and Zhou]{huang2024selfcorrect}
Jie Huang, Xinyun Chen, Swaroop Mishra, Huaixiu~Steven Zheng, Adams~Wei Yu, Xinying Song, and Denny Zhou.
\newblock Large language models cannot self-correct reasoning yet.
\newblock In \emph{International Conference on Learning Representations}, 2024.

\bibitem[Huang et~al.(2026)Huang, Huang, Choubey, Zhou, Chen, May, and Wu]{huang2026gta}
Tenghao Huang, Kung-Hsiang Huang, Prafulla~Kumar Choubey, Yilun Zhou, Muhao Chen, Jonathan May, and Chien-Sheng Wu.
\newblock {GTA}: Generating long-horizon tasks for web agents at scale.
\newblock In \emph{Proceedings of the 64th Annual Meeting of the Association for Computational Linguistics (Volume 1: Long Papers)}, pages 18805--18820, 2026.
\newblock \doi{10.18653/v1/2026.acl-long.856}.

\bibitem[Jia and Harman(2011)]{jia2011mutation}
Yue Jia and Mark Harman.
\newblock An analysis and survey of the development of mutation testing.
\newblock \emph{IEEE Transactions on Software Engineering}, 37\penalty0 (5):\penalty0 649--678, 2011.
\newblock \doi{10.1109/TSE.2010.62}.

\bibitem[Kamoi et~al.(2024)Kamoi, Zhang, Zhang, Han, and Zhang]{kamoi2024selfcorrect}
Ryo Kamoi, Yusen Zhang, Nan Zhang, Jiawei Han, and Rui Zhang.
\newblock When can {LLMs} actually correct their own mistakes? {A} critical survey of self-correction of {LLMs}.
\newblock \emph{Transactions of the Association for Computational Linguistics}, 12:\penalty0 1417--1440, 2024.
\newblock \doi{10.1162/tacl_a_00713}.

\bibitem[Levenshtein(1966)]{levenshtein1966binary}
Vladimir~I. Levenshtein.
\newblock Binary codes capable of correcting deletions, insertions, and reversals.
\newblock \emph{Soviet Physics Doklady}, 10\penalty0 (8):\penalty0 707--710, 1966.

\bibitem[Levy et~al.(2026)Levy, Wiesel, Marreed, Oved, Yaeli, and Shlomov]{levy2026st}
Ido Levy, Ben Wiesel, Sami Marreed, Alon Oved, Avi Yaeli, and Segev Shlomov.
\newblock {ST-WebAgentBench}: A benchmark for evaluating safety and trustworthiness in web agents.
\newblock In \emph{International Conference on Learning Representations}, 2026.

\bibitem[Mesbah et~al.(2012)Mesbah, van Deursen, and Lenselink]{mesbah2012crawljax}
Ali Mesbah, Arie van Deursen, and Stefan Lenselink.
\newblock Crawling {Ajax}-based web applications through dynamic analysis of user interface state changes.
\newblock \emph{ACM Transactions on the Web}, 6\penalty0 (1):\penalty0 3:1--3:30, 2012.
\newblock \doi{10.1145/2109205.2109208}.

\bibitem[{morluto}(2026)]{morluto2026rea}
{morluto}.
\newblock {REA}: Reverse engineer anything.
\newblock GitHub repository, \url{https://github.com/morluto/rea}, 2026.
\newblock Accessed 2026-11-20.

\bibitem[Nam et~al.(2026)Nam, Dennis, Jaques, and Chang]{nam2026evolving}
Hyunji~Alex Nam, Michael Dennis, Natasha Jaques, and Minsuk Chang.
\newblock Evolving self-tuning benchmarks for long-horizon tool use.
\newblock \emph{arXiv preprint arXiv:2610.27307}, 2026.

\bibitem[Qi et~al.(2025)Qi, Liu, Iong, Lai, Sun, Sun, Yang, Yang, Yao, Xu, Tang, and Dong]{ICLR2025_c66e1fcc}
Zehan Qi, Xiao Liu, Iat~Long Iong, Hanyu Lai, Xueqiao Sun, Jiadai Sun, Xinyue Yang, Yu~Yang, Shuntian Yao, Wei Xu, Jie Tang, and Yuxiao Dong.
\newblock {WebRL}: Training {LLM} web agents via self-evolving online curriculum reinforcement learning.
\newblock In \emph{International Conference on Learning Representations}, 2025.

\bibitem[Rawles et~al.(2025)Rawles, Clinckemaillie, Chang, Waltz, Lau, Fair, Li, Bishop, Li, Campbell-Ajala, Toyama, Berry, Tyamagundlu, Lillicrap, and Riva]{rawles2025androidworld}
Christopher Rawles, Sarah Clinckemaillie, Yifan Chang, Jonathan Waltz, Gabrielle Lau, Marybeth Fair, Alice Li, William Bishop, Wei Li, Folawiyo Campbell-Ajala, Daniel Toyama, Robert Berry, Divya Tyamagundlu, Timothy Lillicrap, and Oriana Riva.
\newblock {AndroidWorld}: A dynamic benchmarking environment for autonomous agents.
\newblock In \emph{International Conference on Learning Representations}, 2025.

\bibitem[Sharma et~al.(2005)Sharma, Wu, and Dalal]{sharma2005ciede2000}
Gaurav Sharma, Wencheng Wu, and Edul~N. Dalal.
\newblock The {CIEDE2000} color-difference formula: Implementation notes, supplementary test data, and mathematical observations.
\newblock \emph{Color Research \& Application}, 30\penalty0 (1):\penalty0 21--30, 2005.
\newblock \doi{10.1002/col.20070}.

\bibitem[Shi et~al.(2025)Shi, Ma, Liang, Diao, Ma, and Vosoughi]{shi2025position}
Lin Shi, Chiyu Ma, Wenhua Liang, Xingjian Diao, Weicheng Ma, and Soroush Vosoughi.
\newblock Judging the judges: A systematic study of position bias in {LLM}-as-a-judge.
\newblock In \emph{Proceedings of the 14th International Joint Conference on Natural Language Processing and the 4th Conference of the Asia-Pacific Chapter of the Association for Computational Linguistics}, pages 292--314, 2025.

\bibitem[Si et~al.(2025)Si, Zhang, Li, Yang, Liu, and Yang]{si2025design2code}
Chenglei Si, Yanzhe Zhang, Ryan Li, Zhengyuan Yang, Ruibo Liu, and Diyi Yang.
\newblock {Design2Code}: Benchmarking multimodal code generation for automated front-end engineering.
\newblock In \emph{Proceedings of the 2025 Conference of the Nations of the {A}mericas Chapter of the Association for Computational Linguistics: Human Language Technologies (Volume 1: Long Papers)}, pages 3956--3974, 2025.
\newblock \doi{10.18653/v1/2025.naacl-long.199}.

\bibitem[Wang et~al.(2026)Wang, Zhou, Tong, Kong, Li, Hoi, Qian, and Wang]{wang2026trace2world}
Zezhou Wang, Hanzhang Zhou, Panrong Tong, Quyu Kong, Pengxiang Li, Steven Hoi, Zhuzhong Qian, and Yue Wang.
\newblock {Trace2World}: Reconstructing app worlds from real interaction traces for scalable {GUI} agent training, 2026.
\newblock \url{https://arxiv.org/abs/2610.30792}.

\bibitem[Wang et~al.(2003)Wang, Simoncelli, and Bovik]{wang2003msssim}
Zhou Wang, Eero~P. Simoncelli, and Alan~C. Bovik.
\newblock Multiscale structural similarity for image quality assessment.
\newblock In \emph{Proceedings of the Thirty-Seventh Asilomar Conference on Signals, Systems and Computers}, pages 1398--1402, 2003.
\newblock \doi{10.1109/ACSSC.2003.1292216}.

\bibitem[Wang et~al.(2004)Wang, Bovik, Sheikh, and Simoncelli]{wang2004ssim}
Zhou Wang, Alan~C. Bovik, Hamid~R. Sheikh, and Eero~P. Simoncelli.
\newblock Image quality assessment: From error visibility to structural similarity.
\newblock \emph{IEEE Transactions on Image Processing}, 13\penalty0 (4):\penalty0 600--612, 2004.
\newblock \doi{10.1109/TIP.2003.819861}.

\bibitem[Wei et~al.(2026)Wei, Ma, Zhao, Zhou, Ni, Gan, and Cohan]{wei2026opencomputer}
Jinbiao Wei, Qianran Ma, Yilun Zhao, Xiao Zhou, Kangqi Ni, Guo Gan, and Arman Cohan.
\newblock {OpenComputer}: Verifiable software worlds for computer-use agents.
\newblock \emph{arXiv preprint arXiv:2605.19769}, 2026.

\bibitem[Xie et~al.(2026)Xie, Xu, Zhao, and Song]{xie2026agentsynth}
Jingxu Xie, Dylan Xu, Xuandong Zhao, and Dawn Song.
\newblock {AgentSynth}: Scalable task generation for generalist computer-use agents.
\newblock In \emph{International Conference on Learning Representations}, 2026.

\bibitem[Xu et~al.(2025)Xu, Song, Li, Tang, Jain, Bao, Wang, Zhou, Guo, Cao, et~al.]{xu2026theagentcompany}
Frank~Fangzheng Xu, Yufan Song, Boxuan Li, Yuxuan Tang, Kritanjali Jain, Mengxue Bao, Zora Wang, Xuhui Zhou, Zhitong Guo, Murong Cao, et~al.
\newblock {TheAgentCompany}: Benchmarking {LLM} agents on consequential real world tasks.
\newblock In \emph{Advances in Neural Information Processing Systems}, volume~38, 2025.

\bibitem[Yin et~al.(2026)Yin, Guan, Wang, Cao, Lin, Cheng-Yue, Long, Wong, Dhingra, Wang, and Zhou]{yin2026constructing}
Xunjian Yin, Tianchen Guan, Jinao Wang, Weili Cao, Daisy~Xinlei Lin, Royce Cheng-Yue, Keagan Long, Kyle Wong, Bhuwan Dhingra, Xiangjun Wang, and Shuyan Zhou.
\newblock Constructing challenging browser-use tasks by controlled environment interventions.
\newblock \emph{arXiv preprint arXiv:2609.35814}, 2026.

\bibitem[Yu et~al.(2026)Yu, Wang, Jiang, Li, and Wang]{yu2026medcua}
Jia Yu, Zilong Wang, Xinyang Jiang, Dongsheng Li, and Shuo Wang.
\newblock {MedCUA-Bench}: A screenshot-only benchmark for clinical computer-use agents.
\newblock In \emph{Findings of the Association for Computational Linguistics: EMNLP 2026}, pages 5687--5712. Association for Computational Linguistics, 2026.
\newblock \doi{10.18653/2026.findings-emnlp.271}.

\bibitem[Zhang et~al.(2026)Zhang, Guo, Goh, Hu, Chen, Wang, Gao, and Shou]{zhang2026showuialoha}
Yichun Zhang, Xiangwu Guo, Yauhong Goh, Jessica Hu, Zhiheng Chen, Xin Wang, Difei Gao, and Mike~Zheng Shou.
\newblock {ShowUI-Aloha}: Human-taught {GUI} agent.
\newblock arXiv preprint arXiv:2601.07181, 2026.

\bibitem[Zheng et~al.(2023)Zheng, Chiang, Sheng, Zhuang, Wu, Zhuang, Lin, Li, Li, Xing, Zhang, Gonzalez, and Stoica]{zheng2023judging}
Lianmin Zheng, Wei-Lin Chiang, Ying Sheng, Siyuan Zhuang, Zhanghao Wu, Yonghao Zhuang, Zi~Lin, Zhuohan Li, Dacheng Li, Eric~P. Xing, Hao Zhang, Joseph~E. Gonzalez, and Ion Stoica.
\newblock Judging {LLM}-as-a-judge with {MT-Bench} and {Chatbot Arena}.
\newblock In \emph{Advances in Neural Information Processing Systems}, volume~36, pages 46595--46623, 2023.
\newblock \doi{10.52202/075280-2020}.

\bibitem[Zhou et~al.(2024)Zhou, Xu, Zhu, Zhou, Lo, Sridhar, Cheng, Ou, Bisk, Fried, et~al.]{zhou2024webarena}
Shuyan Zhou, Frank~F. Xu, Hao Zhu, Xuhui Zhou, Robert Lo, Abishek Sridhar, Xianyi Cheng, Tianyue Ou, Yonatan Bisk, Daniel Fried, et~al.
\newblock {WebArena}: A realistic web environment for building autonomous agents.
\newblock In \emph{International Conference on Learning Representations}, 2024.

\bibitem[Zhu et~al.(2025)Zhu, Jin, Pruksachatkun, Zhang, Liu, Cui, Kapoor, Longpre, Meng, Weiss, Barez, Gupta, Dhamala, Merizian, Giulianelli, Coppock, Ududec, Kellermann, Sekhon, Steinhardt, Schwettmann, Narayanan, Zaharia, Stoica, Liang, and Kang]{zhu2025abc}
Yuxuan Zhu, Tengjun Jin, Yada Pruksachatkun, Andy Zhang, Shu Liu, Sasha Cui, Sayash Kapoor, Shayne Longpre, Kevin Meng, Rebecca Weiss, Fazl Barez, Rahul Gupta, Jwala Dhamala, Jacob Merizian, Mario Giulianelli, Harry Coppock, Cozmin Ududec, Antony Kellermann, Jasjeet Sekhon, Jacob Steinhardt, Sarah Schwettmann, Arvind Narayanan, Matei Zaharia, Ion Stoica, Percy Liang, and Daniel Kang.
\newblock Establishing best practices for building rigorous agentic benchmarks.
\newblock In \emph{Advances in Neural Information Processing Systems}, volume~38, pages 184435--184475, 2025.
\newblock \doi{10.52202/085713-5547}.

\end{thebibliography}

\clearpage
\appendix

\section*{Appendix}

This appendix gives the details needed to reproduce and check the comparison in~\Cref{sec:exp}.~\Cref{app:request} gives the construction request and recordings that every build received, and~\Cref{app:judges} gives the settings and prompts of the two LLM judges.~\Cref{app:perbuild} reports each build and each build pair behind~\Cref{tab:pairs,tab:headline}.~\Cref{app:working} describes the working-task test and why tasks fail it.~\Cref{app:tasks} gives a complete \PluginName{} task and a flawed task generated by a baseline build, and~\Cref{app:screens} shows additional screens.

\section{Construction Request and Inputs}\label{app:request}

We give every build the same construction request and recordings, so that the two methods differ only in whether Claude Code runs with \PluginName{}.~\Cref{fig:request} shows the construction request. It is identical for every build except for its port numbers. \PluginName{} builds receive it after the command \mbox{\texttt{/bench-clone:clone-from-videos}}, and baseline builds receive the request alone.

\begin{figure}[!ht]
\centering
\fbox{\parbox{\dimexpr\linewidth-2\fboxsep-2\fboxrule\relax}{\footnotesize\setlength{\parskip}{4pt}
Clone the web app shown in the screen recordings in ./recordings into a benchmark environment in this folder, so computer-use agents can be tested on the work the recordings show. Agents see 1280x720 screenshots and act by clicking and typing. Each task should be a piece of real work from the recordings, done end to end, hard enough that clicking around cannot pass it. Each task needs at least 15 golden actions (scroll steps do not count) and at least 5 evals; an eval may carry \texttt{"points"} (default 1): give the eval on the task's decision or main saved outcome \texttt{"points": 3}.

It is scored automatically, so keep to this interface. \texttt{python3 server.py -{}-port P} serves the clone at \texttt{http://127.0.0.1:P/} (Python standard library only). \texttt{POST /\_\_bench/reset} restores the start state and answers 200 (it may carry a JSON body); \texttt{GET /\_\_bench/state} returns the app state as JSON. Write the tasks yourself, one file each, \texttt{tasks/\textless{}id\textgreater{}.json} with: \texttt{"id"} (= the file name), \texttt{"goal"}, \texttt{"start"} (a path beginning with /), \texttt{"max\_steps"} (at least twice the golden actions), \texttt{"step\_by\_step"} (a list of the steps in plain words), \texttt{"golden"} (the same steps as UI actions a browser replays: \texttt{\{"click": "\textless{}visible text\textgreater{}"\}}, \texttt{\{"fill": "\textless{}field label\textgreater{}", "text": "..."\}}, \texttt{\{"select": "\textless{}field label\textgreater{}", "option": "..."\}}, \texttt{\{"key": "Enter"\}}, \texttt{\{"scroll": \textless{}pixels\textgreater{}\}}; add \texttt{"nth": k} (0-based) when several visible controls share the text; no other keys) and \texttt{"evals"} (at least one \texttt{\{"type": "jmespath", "query": "\textless{}JMESPath over the state\textgreater{}", "expected\_value": \textless{}JSON\textgreater{}, "description": "..."\}} per task, plus \texttt{\{"type": "llm\_judge", "rubric": "..."\}} for free text).

Write tasks the way a hard benchmark does: the goal names the record and the job in a few sentences, as a supervisor would ask it, and never gives the answer, lists the options, or says which screens or buttons to use; the right outcome must follow from facts the agent finds on the screens. Each task's golden path proves the task can be done from the start page. Check the decision (points 3), every change the task makes, and that the likely wrong action and changes to other records did not happen. Every task is different work or a different answer: copies count once. A goal is 3 to 6 sentences and quotes no text for the agent to copy (no quoted run of 6 or more words). Checks that only confirm a screen was viewed carry at most 30\% of a task's points, and a run that gives the usual wrong answer must score under 60\%.

Nobody will answer questions: decide yourself. Nothing can be installed; Python Playwright with Chromium is there for your own checks; only loopback ports [ports] work. [time and resource limits] When you are done, write \texttt{"State: DONE"} in \texttt{STATUS.md}.
}}
\caption{Construction request given to every build. Square brackets mark two omissions: the port numbers and the time and resource limits.}
\label{fig:request}
\end{figure}

\Cref{tab:inputs} lists the inputs of each application. Every build of an application receives one application tour and one or two workflow recordings. The reference screens are the denominator of reach, so~\Cref{tab:headline}, which pools two builds, scores prior authorization out of 16 screens.

\begin{table}[!ht]
\centering
\small
\setlength{\tabcolsep}{7pt}
\renewcommand{\arraystretch}{1.12}
\begin{tabular}{@{}lccc@{}}
\toprule
\textbf{Input} & \textbf{Prior authorization} & \textbf{Appeals and denials} & \textbf{DME} \\
\midrule
Application tour (minutes:seconds) & 4:45 & 4:14 & 3:30 \\
Reference screens & 8 & 2 & 5 \\
Screen clusters & 3 & 2 & 2 \\
\bottomrule
\end{tabular}
\caption{Inputs for each application. Reference screens are the recording frames used for visual preference, visual similarity, text match and reach. Screen clusters group reference screens that differ only in record-specific content, and each cluster has equal weight.}
\label{tab:inputs}
\end{table}

\FloatBarrier
\section{LLM Judges}\label{app:judges}

Two of the six measures rely on an LLM judge: visual preference and task preference.~\Cref{tab:judges} gives the settings of each judge, and~\Cref{fig:prompt-vp,fig:prompt-tp} give their prompts. The visual judge compares one reference screen with one screenshot from each build. The task judge compares the complete task sets of two builds against a HealthAdminBench task. It sees each generated task as its goal, start page, numbered golden path, and checks and rubrics, without the query paths or points.

\begin{table}[!ht]
\centering
\small
\setlength{\tabcolsep}{5pt}
\renewcommand{\arraystretch}{1.12}
\begin{tabular}{@{}lll@{}}
\toprule
\textbf{Setting} & \textbf{Visual preference} & \textbf{Task preference} \\
\midrule
Judge model & Claude Opus 5.5 & Claude Opus 5.5 \\
Reasoning effort & medium & medium \\
Unit compared & reference screen & recorded workflow \\
Judge input & reference screen and two screenshots & HealthAdminBench task and two task sets \\
Answers allowed & X or Y & X or Y \\
Samples per presentation order & 3 & 3 \\
\bottomrule
\end{tabular}
\caption{Settings of the two LLM judges. Each comparison runs in both presentation orders, and a JSON schema fixes the answer format.}
\label{tab:judges}
\end{table}

\begin{figure}[!ht]
\centering
\fbox{\parbox{\dimexpr\linewidth-2\fboxsep-2\fboxrule\relax}{\footnotesize\setlength{\parskip}{4pt}
Image R is a frame from a screen recording of a web application (ignore any red click circle or cursor). Images X and Y are screenshots of two rebuilt copies of that application, each showing the state that copy offers for the screen in R. Which screenshot is CLOSER to R as a user would see it, in shell (header, navigation), layout (regions and their arrangement), components (tabs, buttons, tables and columns, form fields) and interface text (headings, labels, tab, button and column texts)? Answer X or Y.}}
\caption{Prompt of the visual preference judge. Images R, X and Y are attached to the prompt.}
\label{fig:prompt-vp}
\end{figure}

\begin{figure}[!ht]
\centering
\fbox{\parbox{\dimexpr\linewidth-2\fboxsep-2\fboxrule\relax}{\footnotesize\setlength{\parskip}{4pt}
You grade benchmark tasks written by a builder who cloned a healthcare admin web app from screen recordings. An expert task (the REFERENCE) describes the work the recordings show. The builder's clone uses its own invented data and its own screen and field names, so names, ids, numbers, dates and labels differ. Match on MEANING (what the agent must do, decide and leave behind), never on values or wording.

REFERENCE (\{tier\}, \{n\_ref\} steps)\\
\{ref\}

\medskip
Two task SETS (X and Y) were written for clones of the app the reference comes from. Each set is the whole set of tasks one builder wrote; values and names differ, so never compare values, wording or the number of tasks. Pick the set that, as a BENCHMARK of the reference's recorded work, is better on these criteria:\\
\hspace*{1em}(a) some task covers the whole recorded workflow end to end (same kind of record, decision or request, and side effects), not just a part of it\\
\hspace*{1em}(b) the decision the recording shows is left to the agent (the goal does not reveal the answer)\\
\hspace*{1em}(c) there is a graded check for every recorded state change and for the decision, with the result the recording shows (checks of a different result count against the set)\\
\hspace*{1em}(e) free-text outcomes (notes, reasons) are graded by a rubric\\
Extra tasks, repeated checks or longer text earn nothing by themselves. Answer X or Y.

SET X:\\
\{x\}

SET Y:\\
\{y\}}}
\caption{Prompt of the task preference judge. The reference block (top) is followed by the comparison prompt (bottom). Braces mark text filled in for each comparison.}
\label{fig:prompt-tp}
\end{figure}

\paragraph{Agreement across presentation orders.} Position bias would appear as a majority answer that changes when the two builds swap places. With Claude Opus 5.5 builds, this happens in none of 57 screen comparisons and in 1 of 16 workflow comparisons. With Claude Sonnet 5.5 builds, it happens in none of 60 screen comparisons and in 2 of 16 workflow comparisons, which the protocol counts as ties. With Claude Haiku 5.5 builds, it happens in 1 of 58 screen comparisons and in none of 16 workflow comparisons, and the protocol counts it as a tie. Most answers favor one method: with Claude Opus 5.5 builds, 329 of 342 visual preference answers and 92 of 96 task preference answers favor the \PluginName{} build. With Claude Sonnet 5.5 builds, 360 of 360 visual preference answers and 76 of 96 task preference answers do. With Claude Haiku 5.5 builds, 327 of 348 visual preference answers and 95 of 96 task preference answers do.

\FloatBarrier
\section{Per-Build and Per-Pair Results}\label{app:perbuild}

\Cref{tab:pairs,tab:headline} aggregate the builds. Here we report each build and each build pair, so that every count in those tables can be traced to individual builds.

\begin{table}[!ht]
\centering
\small
\setlength{\tabcolsep}{4pt}
\renewcommand{\arraystretch}{1.2}
\begin{tabular*}{\textwidth}{@{\extracolsep{\fill}}lcccccccccccc@{}}
\toprule
& \multicolumn{4}{c}{\textbf{Prior authorization}} & \multicolumn{4}{c}{\textbf{Appeals and denials}} & \multicolumn{4}{c}{\textbf{DME}} \\
\cmidrule(lr){2-5}\cmidrule(lr){6-9}\cmidrule(lr){10-13}
& \multicolumn{2}{c}{\textbf{Ours}} & \multicolumn{2}{c}{\textbf{Plain}} & \multicolumn{2}{c}{\textbf{Ours}} & \multicolumn{2}{c}{\textbf{Plain}} & \multicolumn{2}{c}{\textbf{Ours}} & \multicolumn{2}{c}{\textbf{Plain}} \\
\cmidrule(lr){2-3}\cmidrule(lr){4-5}\cmidrule(lr){6-7}\cmidrule(lr){8-9}\cmidrule(lr){10-11}\cmidrule(lr){12-13}
\textbf{Measure} & \textbf{1} & \textbf{2} & \textbf{1} & \textbf{2} & \textbf{1} & \textbf{2} & \textbf{1} & \textbf{2} & \textbf{1} & \textbf{2} & \textbf{1} & \textbf{2} \\
\midrule
\multicolumn{13}{@{}l}{\textit{Claude Opus 5.5}} \\
Visual similarity & 0.88 & 0.59 & 0.48 & 0.42 & 0.90 & 0.89 & 0.69 & 0.62 & 0.86 & 0.83 & 0.66 & 0.60 \\
Text match & 0.91 & 0.59 & 0.35 & 0.34 & 0.88 & 0.88 & 0.73 & 0.45 & 0.78 & 0.77 & 0.70 & 0.58 \\
Reach & 7/8 & 6/8 & 5/8 & 6/8 & 2/2 & 2/2 & 2/2 & 1/2 & 5/5 & 5/5 & 5/5 & 4/5 \\
Working tasks & 8/10 & 5/5 & 1/7 & 1/7 & 7/8 & 5/5 & 4/9 & 0/10 & 7/7 & 5/5 & 0/6 & 0/7 \\
Check strength & 56/56 & 24/24 & 47/59 & 47/55 & 22/22 & 11/11 & 57/70 & 39/54 & 23/23 & 14/14 & 2/3 & 8/13 \\
\midrule
\multicolumn{13}{@{}l}{\textit{Claude Sonnet 5.5}} \\
Visual similarity & 0.79 & 0.83 & 0.42 & 0.38 & 0.88 & 0.86 & 0.50 & 0.52 & 0.81 & 0.82 & 0.49 & 0.54 \\
Text match & 0.91 & 0.88 & 0.34 & 0.28 & 0.78 & 0.86 & 0.29 & 0.24 & 0.78 & 0.72 & 0.18 & 0.22 \\
Reach & 7/8 & 7/8 & 7/8 & 6/8 & 2/2 & 2/2 & 1/2 & 2/2 & 5/5 & 5/5 & 5/5 & 3/5 \\
Working tasks & 8/8 & 5/8 & 2/4 & 0/4 & 6/6 & 6/6 & 4/5 & 0/6 & 4/4 & 1/5 & 0/5 & 0/5 \\
Check strength & 55/55 & 12/12 & 16/16 & 0/0 & 11/11 & 6/6 & 21/22 & 0/0 & 15/15 & 5/5 & 0/0 & 17/22 \\
\midrule
\multicolumn{13}{@{}l}{\textit{Claude Haiku 5.5}} \\
Visual similarity & 0.55 & 0.75 & 0.39 & 0.35 & 0.78 & 0.76 & 0.40 & 0.55 & 0.76 & 0.77 & 0.47 & 0.47 \\
Text match & 0.51 & 0.78 & 0.38 & 0.24 & 0.71 & 0.59 & 0.21 & 0.18 & 0.75 & 0.75 & 0.21 & 0.36 \\
Reach & 6/8 & 7/8 & 8/8 & 7/8 & 2/2 & 1/2 & 1/2 & 0/2 & 5/5 & 5/5 & 4/5 & 3/5 \\
Working tasks & 5/5 & 5/5 & 0/4 & 1/4 & 3/3 & 4/4 & 4/4 & 0/4 & 4/4 & 3/3 & 0/4 & 2/4 \\
Check strength & 22/22 & 32/32 & 18/26 & 22/27 & 3/3 & 4/4 & 8/8 & 14/14 & 13/13 & 10/10 & 5/6 & 6/6 \\
\bottomrule
\end{tabular*}
\caption{Performance of each build of \PluginName{} (Ours) and Claude Code alone (Plain). Columns 1 and 2 are the two builds of each method. For Claude Opus 5.5,~\Cref{tab:headline} averages the unrounded visual similarity and text match of the two builds and pools the counts. Reach counts reached reference screens, working tasks counts working tasks out of generated tasks, and check strength counts verified values out of applicable entered values.}
\label{tab:perbuild}
\end{table}

Most differences in~\Cref{tab:headline} also hold between individual builds (\Cref{tab:perbuild}). On text match and visual similarity, every \PluginName{} build scores above both baseline builds of its application, with every base model. On check strength, every \PluginName{} build verifies all of its applicable entered values. With Claude Opus 5.5, all six baseline builds leave some entered values unchecked. With Claude Sonnet 5.5, baseline Build 1 for appeals and denials verifies 21 of 22, and baseline Build 2 for DME verifies 17 of 22. With Claude Haiku 5.5, the baseline builds for prior authorization verify 18 of 26 and 22 of 27, and baseline Build 1 for DME verifies 5 of 6.

\begin{table}[!ht]
\centering
\small
\setlength{\tabcolsep}{4pt}
\renewcommand{\arraystretch}{1.2}
\begin{tabular*}{\textwidth}{@{\extracolsep{\fill}}lcccccccccccc@{}}
\toprule
& \multicolumn{4}{c}{\textbf{Prior authorization}} & \multicolumn{4}{c}{\textbf{Appeals and denials}} & \multicolumn{4}{c}{\textbf{DME}} \\
\cmidrule(lr){2-5}\cmidrule(lr){6-9}\cmidrule(lr){10-13}
& \multicolumn{2}{c}{\textbf{Ours 1}} & \multicolumn{2}{c}{\textbf{Ours 2}} & \multicolumn{2}{c}{\textbf{Ours 1}} & \multicolumn{2}{c}{\textbf{Ours 2}} & \multicolumn{2}{c}{\textbf{Ours 1}} & \multicolumn{2}{c}{\textbf{Ours 2}} \\
\cmidrule(lr){2-3}\cmidrule(lr){4-5}\cmidrule(lr){6-7}\cmidrule(lr){8-9}\cmidrule(lr){10-11}\cmidrule(lr){12-13}
\textbf{Measure} & \textbf{P1} & \textbf{P2} & \textbf{P1} & \textbf{P2} & \textbf{P1} & \textbf{P2} & \textbf{P1} & \textbf{P2} & \textbf{P1} & \textbf{P2} & \textbf{P1} & \textbf{P2} \\
\midrule
\multicolumn{13}{@{}l}{\textit{Claude Opus 5.5}} \\
Visual preference & W & W & W & W & W & W & W & W & W & W & W & W \\
Visual similarity & W & W & W & W & W & W & W & W & W & W & W & W \\
Text match & W & W & W & W & W & W & W & W & W & W & W & W \\
Task preference & W & W & W & W & W & W & W & W & T & W & W & W \\
Working tasks & W & W & W & W & W & W & W & W & W & W & W & W \\
Check strength & W & W & W & W & W & W & W & W & W & W & W & W \\
\textbf{Wins} & \textbf{6} & \textbf{6} & \textbf{6} & \textbf{6} & \textbf{6} & \textbf{6} & \textbf{6} & \textbf{6} & \textbf{5} & \textbf{6} & \textbf{6} & \textbf{6} \\
\midrule
\multicolumn{13}{@{}l}{\textit{Claude Sonnet 5.5}} \\
Visual preference & W & W & W & W & W & W & W & W & W & W & W & W \\
Visual similarity & W & W & W & W & W & W & W & W & W & W & W & W \\
Text match & W & W & W & W & W & W & W & W & W & W & W & W \\
Task preference & W & W & W & W & W & W & L & L & W & W & T & W \\
Working tasks & W & W & W & W & W & W & W & W & W & W & W & W \\
Check strength & T & W & T & W & W & W & W & W & W & W & W & W \\
\textbf{Wins} & \textbf{5} & \textbf{6} & \textbf{5} & \textbf{6} & \textbf{6} & \textbf{6} & \textbf{5} & \textbf{5} & \textbf{6} & \textbf{6} & \textbf{5} & \textbf{6} \\
\midrule
\multicolumn{13}{@{}l}{\textit{Claude Haiku 5.5}} \\
Visual preference & W & W & W & W & W & W & W & W & W & W & W & W \\
Visual similarity & W & W & W & W & W & W & W & W & W & W & W & W \\
Text match & W & W & W & W & W & W & W & W & W & W & W & W \\
Task preference & W & W & W & W & W & W & W & W & W & W & W & W \\
Working tasks & W & W & W & W & L & W & T & W & W & W & W & W \\
Check strength & W & W & W & W & T & T & T & T & W & T & W & T \\
\textbf{Wins} & \textbf{6} & \textbf{6} & \textbf{6} & \textbf{6} & \textbf{4} & \textbf{5} & \textbf{4} & \textbf{5} & \textbf{6} & \textbf{5} & \textbf{6} & \textbf{5} \\
\bottomrule
\end{tabular*}
\caption{Outcome of each build pair for \PluginName{} (Ours) against Claude Code alone (Plain). Each column pairs one Ours build with Plain Build 1 (P1) or Plain Build 2 (P2) of the same application. W, T and L mark a win, tie or loss for \PluginName{}. The Wins row counts wins out of six measures, and its totals are the wins in~\Cref{tab:pairs}.}
\label{tab:perpair}
\end{table}

With Claude Opus 5.5, \PluginName{} wins all six measures in 11 of the 12 pairs (\Cref{tab:perpair}). With Claude Sonnet 5.5, it wins all six measures in 7 of the 12 pairs. Its five other outcomes are the two task preference losses for appeals and denials (\Cref{sec:results}), a task preference tie for DME, and two check-strength ties with baseline Build 1 for prior authorization, because each build verifies every applicable value. With Claude Haiku 5.5, it wins all six measures in 6 of the 12 pairs. The eight other outcomes are:
\begin{itemize}[leftmargin=*,itemsep=1pt]
    \item Six ties on check strength, in all four pairs for appeals and denials and in both DME pairs with baseline Build 2, because each build verifies every applicable value.
    \item For appeals and denials, baseline Build 1 wins working tasks against \PluginName{} Build 1, with 4 working tasks to its 3, and ties \PluginName{} Build 2, which also has 4.
\end{itemize}

With Claude Opus 5.5, every preference share is 1.00 except in five comparisons. With Claude Sonnet 5.5, every preference share is 1.00 except in four task preference comparisons. The share is 0.75 for prior authorization, 0.00 in both losses, and 0.50 in the tie. With Claude Haiku 5.5, every preference share is 1.00 except in four visual preference comparisons, all for prior authorization. \PluginName{} Build 1 has a share of 0.67 against both baseline builds, and \PluginName{} Build 2 has shares of 0.89 and 0.92.

With Claude Opus 5.5, no check-strength win comes from the rule for builds without applicable entered values. With Claude Sonnet 5.5, six of the ten check-strength wins come from it, because baseline Build 2 for prior authorization, Build 2 for appeals and denials, and Build 1 for DME generate tasks but no applicable entered values. With Claude Haiku 5.5, none of the check-strength wins come from it. No win on working tasks or check strength changes under the sensitivity checks of~\Cref{sec:exp}.

\FloatBarrier
\section{Working-Task Test}\label{app:working}

A task counts as working when an agent can complete it and its validator scores the result correctly. The scoring script tests this by resetting the environment and replaying the task in Chromium at 1280$\times$720 pixels. A task is working only if it passes all six tests below.
\begin{itemize}[leftmargin=*,itemsep=1pt]
    \item \textbf{Format.} The task file has every field the construction request specifies.
    \item \textbf{Reset.} The task starts unsolved, and the reset state is identical before and after the golden path.
    \item \textbf{Golden path.} Each action resolves to one visible, enabled and uncovered control, and every state-based check holds at the end.
    \item \textbf{Simulated incorrect executions.} The state-based checks reject the golden path without its last state-changing action, without any single action, with other text of the same form in a typed field, with the next option in a selection, and with the next of several controls that share a label. Executions that end in the golden path's final state are excluded, and free text graded by a rubric is exempt from the other-text executions. They also reject the final state of every other task's golden path. Rubrics are not run in this test.
    \item \textbf{Correct variations.} The validator accepts the golden path with independent clicks reordered and with each typed text of four or more words reduced to words from the goal and initial data.
    \item \textbf{Required input values.} Every value the golden path enters appears in the goal or the initial data.
\end{itemize}

\begin{table}[!ht]
\centering
\small
\setlength{\tabcolsep}{7pt}
\renewcommand{\arraystretch}{1.12}
\begin{tabular}{@{}lrr@{}}
\toprule
\textbf{Measure} & \textbf{Ours} & \textbf{Plain} \\
\midrule
\multicolumn{3}{@{}l}{\textit{Tasks}} \\
Generated & 40 & 46 \\
Working & 37 & 6 \\
Not working & 3 & 40 \\
\midrule
\multicolumn{3}{@{}l}{\textit{Reasons a task is not working}} \\
Reset fails or task starts solved & 0 & 0 \\
Golden path fails to replay or pass its checks & 0 & 8 \\
Incomplete execution accepted & 0 & 27 \\
Wrong typed or selected value accepted & 0 & 29 \\
Wrong control with the same label accepted & 0 & 6 \\
Correct variation rejected & 0 & 1 \\
Input value missing from goal and initial data & 3 & 3 \\
\bottomrule
\end{tabular}
\caption{Working-task test results with Claude Opus 5.5 for the six builds of \PluginName{} (Ours) and the six builds of Claude Code alone (Plain). A task can fail for more than one reason.}
\label{tab:working}
\end{table}

With Claude Opus 5.5, 3 of the 40 \PluginName{} tasks and 40 of the 46 baseline tasks are not working (\Cref{tab:working}). For the baseline, 32 of the 40 tasks that are not working have state-based checks that accept a wrong execution: 27 accept an incomplete execution, and 29 accept a wrong typed or selected value. These baseline tasks therefore fail mainly because their state-based checks do not detect a wrong result. With Claude Sonnet 5.5, 7 of the 37 \PluginName{} tasks and 23 of the 29 baseline tasks are not working, most of them because their golden path fails to replay or to pass its own checks. With Claude Haiku 5.5, every \PluginName{} task passes the test, and 17 of the 24 baseline tasks are not working. No task of either method fails the reset test.

\FloatBarrier
\needspace{0.8\textheight}
\section{Example Tasks}\label{app:tasks}

\Cref{fig:task-s2b,fig:task-base} show two tasks for the same recorded DME workflow: fax the equipment supplier the documents it requires, record the fax in a signed progress note, and clear the patient from the worklist. The first task was generated by a \PluginName{} build and the second by a baseline build, and both goals leave the document set to the agent. Each figure shows the screen that matters most for the task, the parts of the golden path that each check covers, and the simulated incorrect executions of the working-task test (\Cref{app:working}). The \PluginName{} task checks every part of the workflow and rejects all 27 of these executions. The baseline task has no state-based check on the progress note, so an execution that never writes or signs the note passes every state-based check, and the task fails the working-task test.

\begin{center}\begin{minipage}{\linewidth}
\captionsetup{type=figure}
\includegraphics[width=\linewidth]{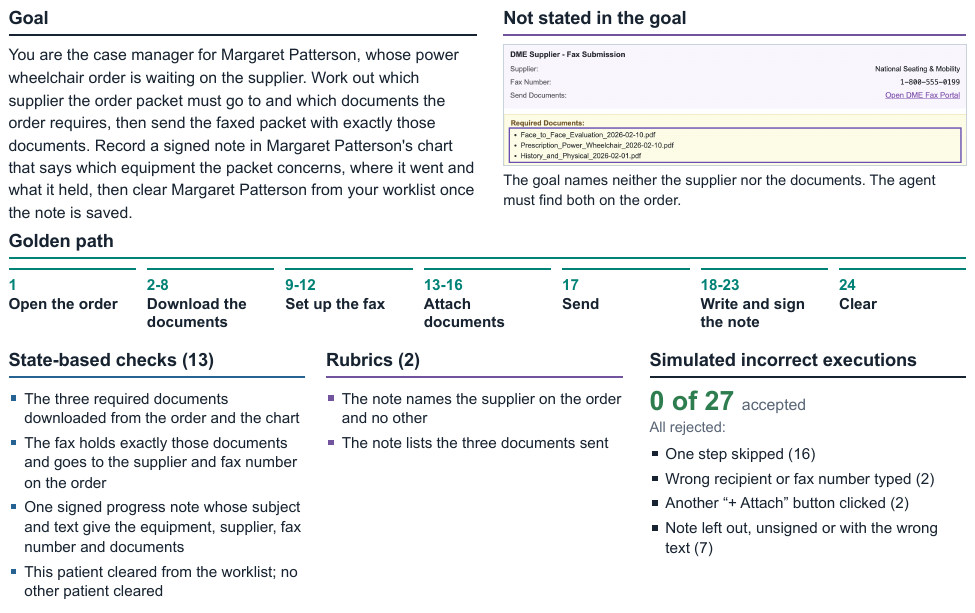}
\caption{Complete evaluation task generated by \PluginName{} for DME (Claude Haiku 5.5, Build 1). The screenshot shows the order, where the agent must find the three documents the supplier requires. The task passes the working-task test, and its validator rejects all 27 simulated incorrect executions.}
\label{fig:task-s2b}
\end{minipage}\end{center}

\begin{center}\begin{minipage}{\linewidth}
\captionsetup{type=figure}
\includegraphics[width=\linewidth]{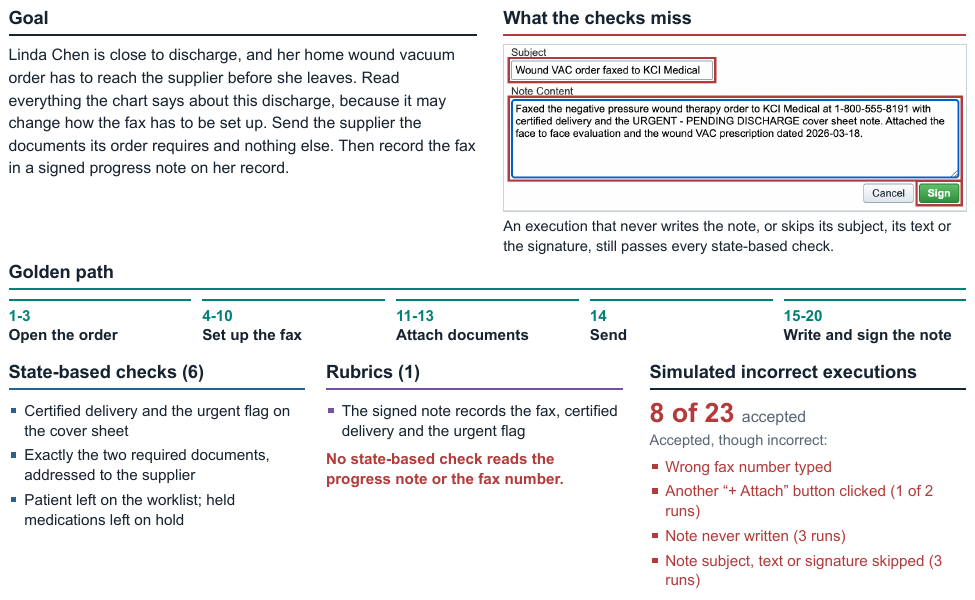}
\caption{Evaluation task generated by a baseline build for the same workflow as~\Cref{fig:task-s2b} (DME, Claude Haiku 5.5, Build 1). The screenshot shows the progress note, which only the rubric covers, so the executions that skip the note subject, the note text or the signature pass every state-based check. The validator accepts 8 of the 23 simulated incorrect executions that the test counts, and the task fails the working-task test.}
\label{fig:task-base}
\end{minipage}\end{center}

\section{Additional Screens}
\label{app:screens}
\begin{figure*}[p]
\centering
\newcommand{\apscreen}[1]{\IfFileExists{figures/fig3_#1.png}{\includegraphics[width=0.47\linewidth,height=0.2548\linewidth,keepaspectratio]{figures/fig3_#1.png}}{\fbox{\parbox[c][\dimexpr0.2548\linewidth-2\fboxsep-2\fboxrule\relax][c]{\dimexpr0.47\linewidth-2\fboxsep-2\fboxrule\relax}{\centering\scriptsize\detokenize{fig3_#1.png}}}}}
\setlength{\tabcolsep}{2pt}
\renewcommand{\arraystretch}{1.2}
\begin{tabular}{@{}cc@{}}
\multicolumn{2}{c}{\textbf{Appeals and denials}} \\ \noalign{\vskip 2pt}
\multicolumn{2}{c}{\small Original} \\
\multicolumn{2}{c}{\apscreen{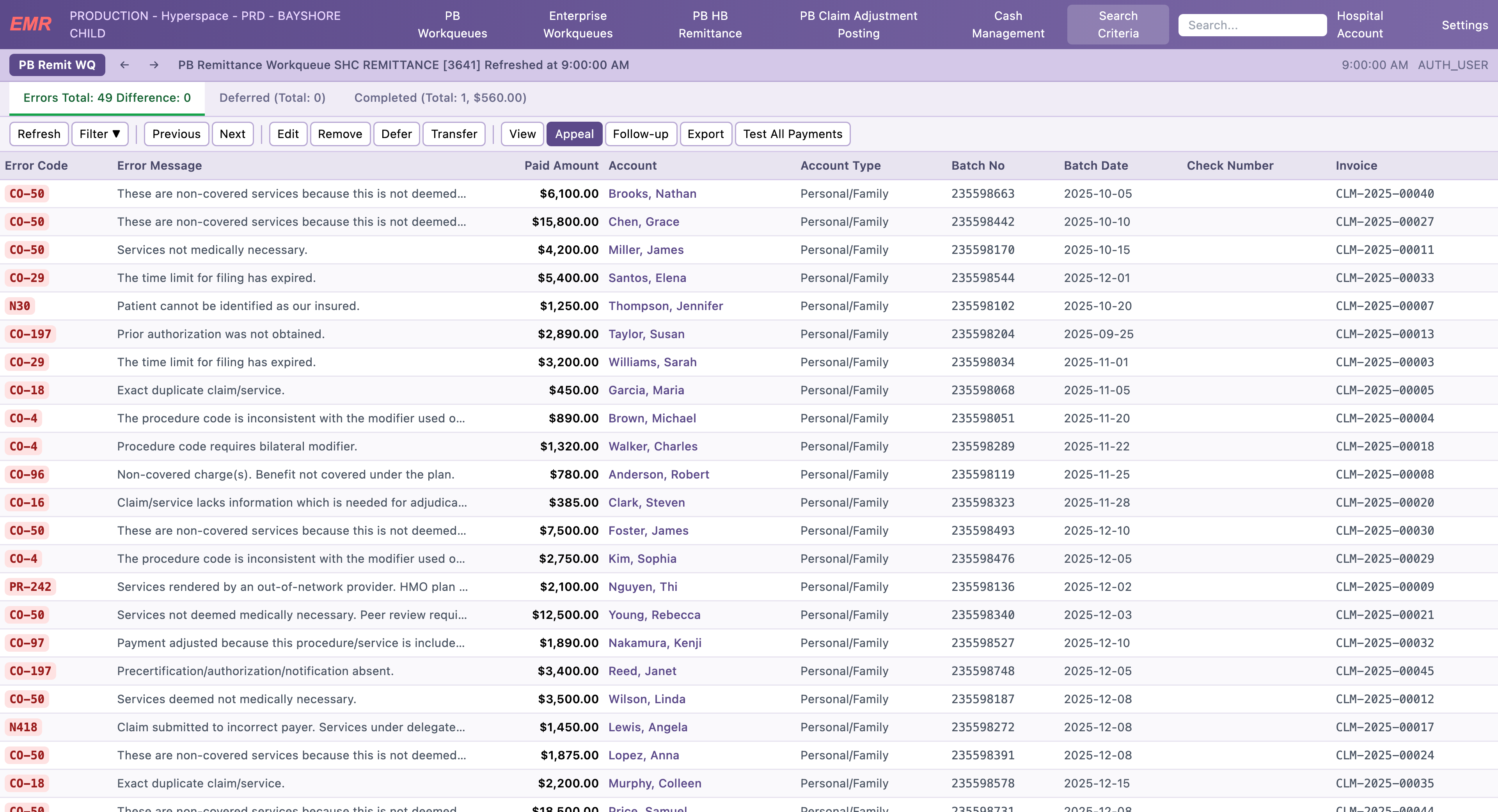}} \\
\small Plain Claude Code & \small \PluginName{} \\
\apscreen{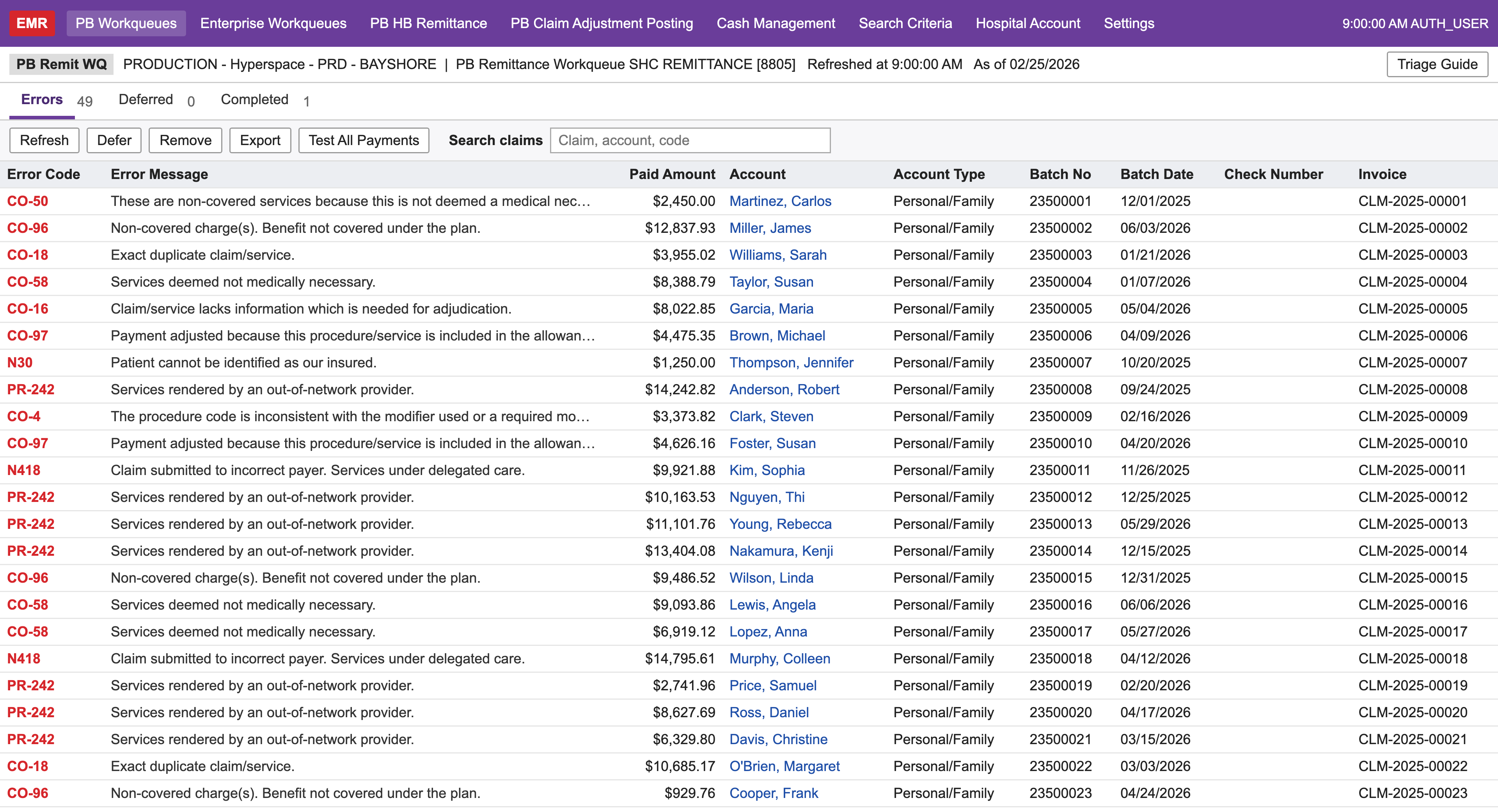} & \apscreen{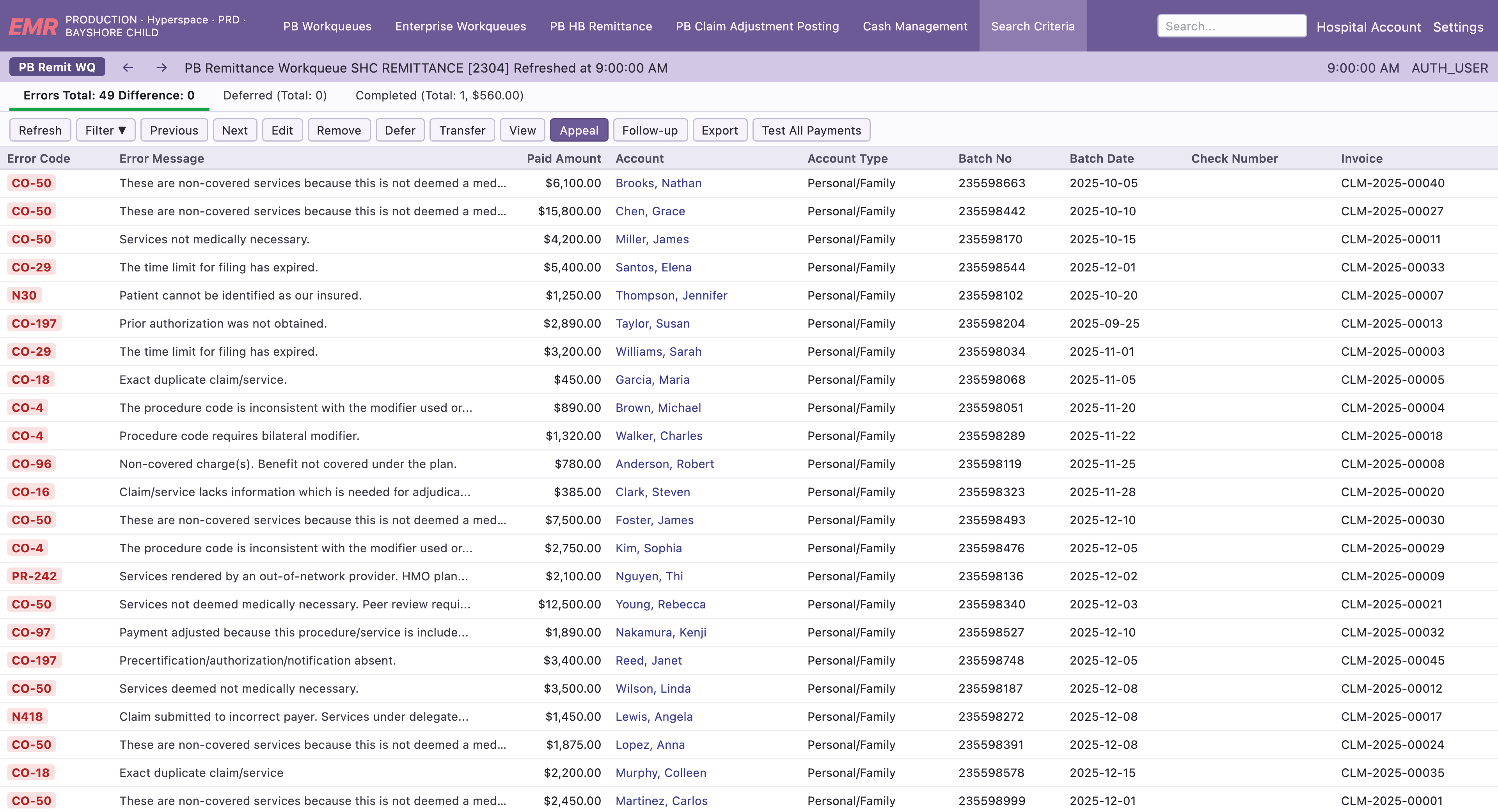} \\ \noalign{\vskip 12pt}
\multicolumn{2}{c}{\textbf{DME}} \\ \noalign{\vskip 2pt}
\multicolumn{2}{c}{\small Original} \\
\multicolumn{2}{c}{\apscreen{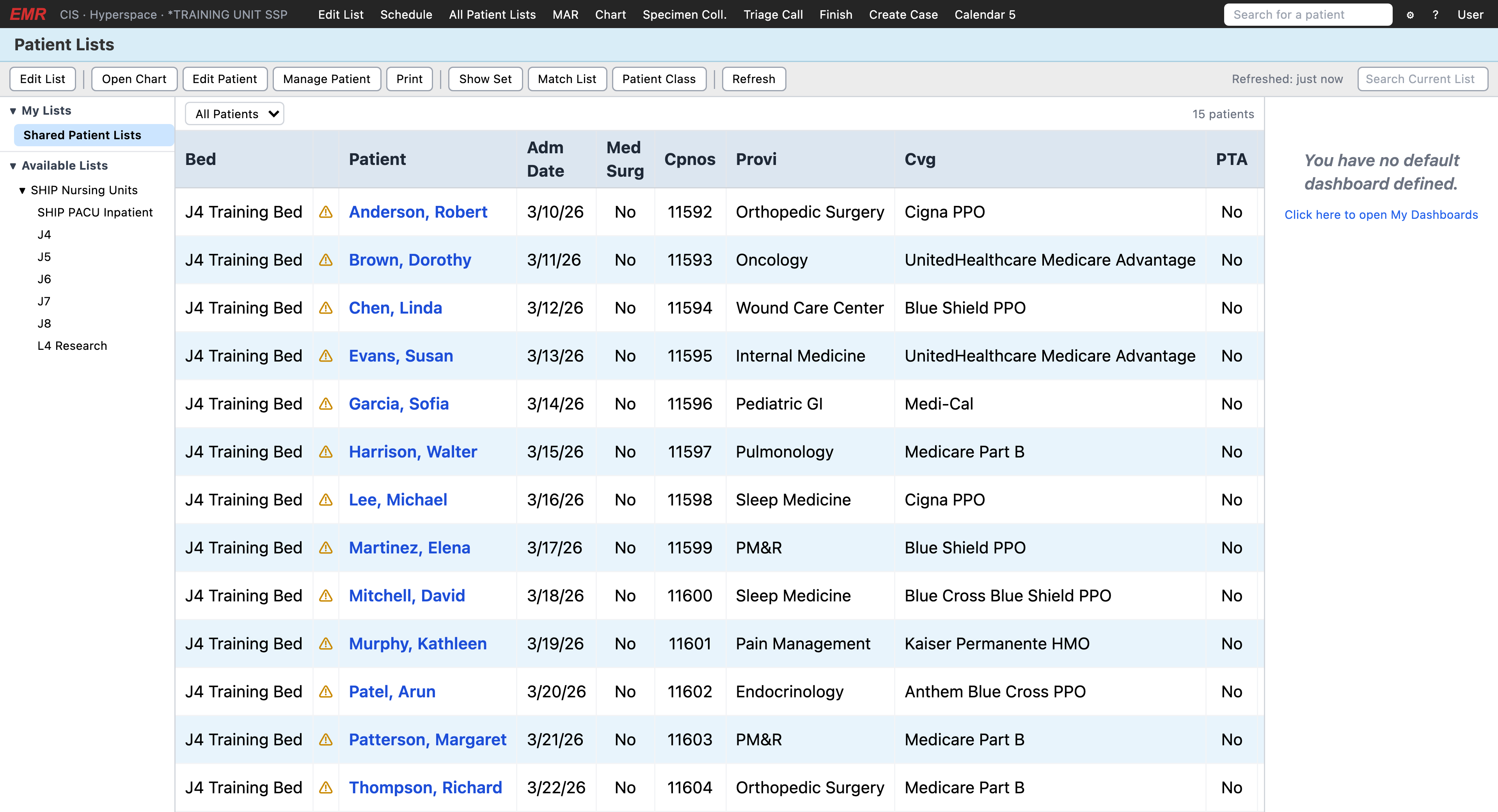}} \\
\small Plain Claude Code & \small \PluginName{} \\
\apscreen{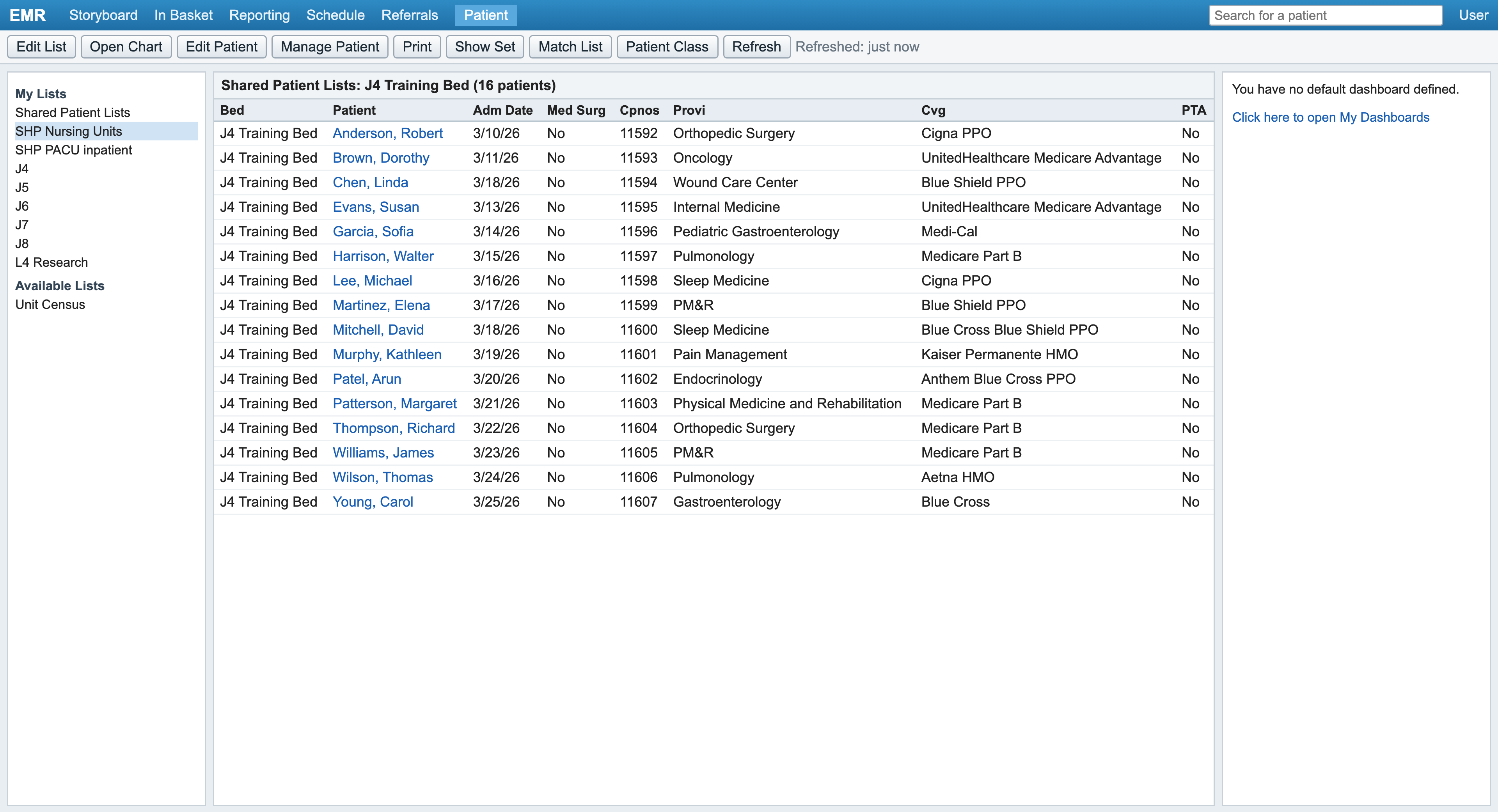} & \apscreen{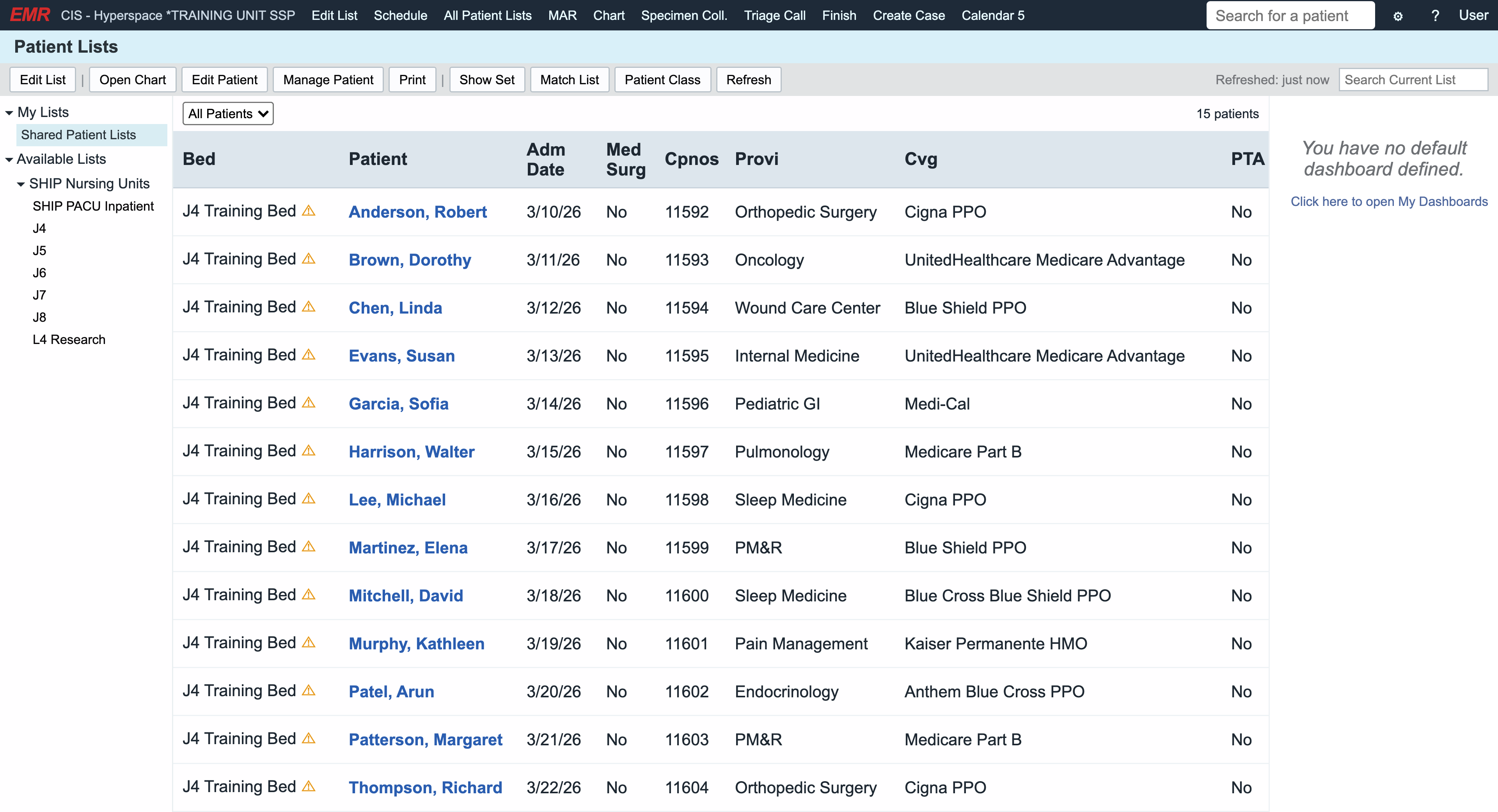} \\
\end{tabular}
\caption{The remittance work list of appeals and denials (upper group) and the patient list of DME (lower group). As in~\Cref{fig:fidelity}, each group shows the original application (top), a baseline build (bottom left) and a \PluginName{} build (bottom right). Appeals and denials shows the second baseline build and the first \PluginName{} build, and DME shows the first build of each. On each screen, the judge preferred the \PluginName{} build to both baseline builds.}
\label{fig:fidelity-more}
\end{figure*}

\Cref{fig:fidelity-more} shows screens of the other two applications in the format of~\Cref{fig:fidelity}.

\end{document}